\documentclass[aps,prx,amsmath,groupedaddress,superscriptaddress,floatfix,twocolumn,showkeys,amsart]{revtex4-1}

\usepackage{multirow}
\usepackage{graphicx} 
\usepackage{dcolumn} 
\usepackage{bm} 
\usepackage{siunitx}
\usepackage{upgreek}
\usepackage{ulem}
\usepackage{xcolor}
\usepackage[pdftex, breaklinks,colorlinks, citecolor=blue, urlcolor=blue]{hyperref}
\usepackage{multirow}
\usepackage{braket} 
\usepackage{amsmath}
\usepackage{comment}
\usepackage{enumitem}
\usepackage{gensymb}
\usepackage{amssymb}

\begin{document}

\title{Time-of-Flight Quantum Tomography of Single Atom Motion}

\author{M.~O.~Brown}
\affiliation{JILA, National Institute of Standards and Technology and University of Colorado, Boulder, CO 80309}
\affiliation{Department of Physics, University of Colorado, Boulder, CO 80309}
\author{S.~R.~Muleady}
\affiliation{JILA, National Institute of Standards and Technology and University of Colorado, Boulder, CO 80309}
\affiliation{Department of Physics, University of Colorado, Boulder, CO 80309}
\affiliation{Center for Theory of Quantum Matter, University of Colorado, Boulder, CO 80309}
\author{W.~J.~Dworschack}
\affiliation{JILA, National Institute of Standards and Technology and University of Colorado, Boulder, CO 80309}
\affiliation{Department of Physics, University of Colorado, Boulder, CO 80309}
\author{R.~J.~Lewis-Swan}
\affiliation{Homer L Dodge Dept of Physics and Astronomy, University of Oklahoma, Norman, OK 73019}
\affiliation{Center for Quantum Research and Technology, University of Oklahoma, Norman, OK 73019}
\author{A.~M.~Rey}
\affiliation{JILA, National Institute of Standards and Technology and University of Colorado, Boulder, CO 80309}
\affiliation{Department of Physics, University of Colorado, Boulder, CO 80309}
\affiliation{Center for Theory of Quantum Matter, University of Colorado, Boulder, CO 80309}
\author{O.~Romero-Isart}
\affiliation{Institute for Theoretical Physics, University of Innsbruck, A-6020 Innsbruck, Austria}
\affiliation{Institute for Quantum Optics and Quantum Information of the Austrian Academy of Sciences, 6020 Innsbruck, Austria}
\author{C.~A.~Regal}
\affiliation{JILA, National Institute of Standards and Technology and University of Colorado, Boulder, CO 80309}
\affiliation{Department of Physics, University of Colorado, Boulder, CO 80309}
\date{March 4, 2022}

\begin{abstract}
Time of flight is an intuitive way to determine the velocity of particles and lies at the heart of many capabilities ranging from mass spectrometry to fluid flow measurements. Here we show time-of-flight imaging can realize tomography of a quantum state of motion of a single trapped atom.  Tomography of motion requires studying the phase space spanned by both position and momentum.  By combining time-of-flight imaging with coherent evolution of the atom in an optical tweezer trap, we are able to access arbitrary quadratures in phase space without relying on coupling to a spin degree of freedom.  To create non-classical motional states, we harness quantum tunneling in the versatile potential landscape of optical tweezers, and our tomography both demonstrates Wigner function negativity and assesses coherence of non-stationary states.  Our demonstrated tomography concept has wide applicability to a range of particles and will enable characterization of non-classical states of more complex systems or massive dielectric particles.

\end{abstract}

\maketitle
The creation and full reconstruction of quantum states featuring genuine non-classical behavior has played a key role in the development of quantum systems. Such reconstructions are perhaps most familiar in quantum optics, where preparing and measuring modes of the electromagnetic field in non-classical states were striking demonstrations of the quantum nature of light. In these experiments, state characterization has been accomplished both with homodyne tomography~\cite{vogel1989determination,smithey1993measurement,lvovsky2001quantum,gross2011atomic} and by coupling photons to a spin degree of freedom in cavity or circuit quantum electrodynamics (QED) ~\cite{deleglise2008reconstruction,hofheinz2009synthesizing}. The associated quasiprobability distributions that are obtained, such as the Wigner function, are useful tools in analyzing non-classical behavior. While quasi-classical coherent states have strictly positive Wigner functions, other states, such as excited Fock states and Schr{\"o}dinger cat states, can exhibit regions of negative phase-space density that have no classical analog.

For particles with mass, the observation of non-classical states of motion is equally intriguing.  Early experiments with trapped ions created non-classical states using trap displacements and Raman sideband transitions, the spin-phonon analog to cavity QED. They verified the generation of squeezed states, Fock states, and cat states among others~\cite{leibfried1996experimental}, and have continued to explore a rich space of tomography methods~\cite{fluhmann2020direct}. Meanwhile, the control of quantum motion of objects has expanded greatly in recent years, for instance with the ability to couple artificial spins to mechanical solid-state acoustical excitations~\cite{oconnell2010quantum,chu2018creation}.  Another example is particle levitation, where one can achieve environmentally-isolated masses whose wavefunction can be expanded over large scales for fundamental studies with massive particles~\cite{chang2010cavity,romero-isart2010toward,gonzalez2021levitodynamics}. It is now possible to cool the center-of-mass motion of a dielectric particle to its quantum ground state~\cite{delic2020cooling,tebbenjohanns2021quantum,magrini2021real}, and it has been proposed that quantum state creation and characterization for these masses can be explored using nonlinear potential landscapes and time-of-flight tomography~\cite{romero-isart2011optically,vanner2013cooling}.  But to date directly measuring rotated quadratures in position and momentum~\cite{dunn1995experimental}, the natural analogy to optical homodyne tomography, has not been harnessed to characterize a non-classical state of a single trapped particle.

\begin{figure*}[tbh]\centering
\includegraphics[width=\textwidth]{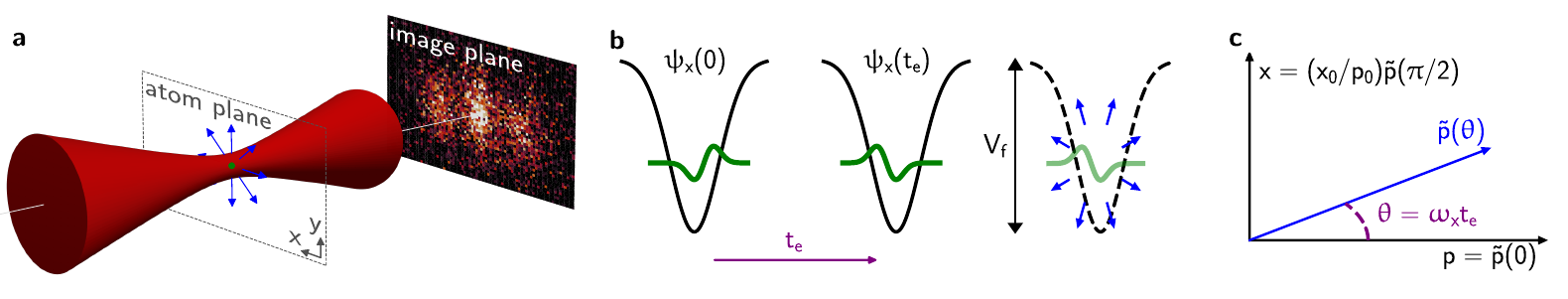}
\caption[Tomography sequence and notation.]{\label{fig:Figure_1} Tomography sequence and notation. \textbf{(a)} Many time-of-flight images of a single atom released from the optical tweezer are averaged to measure the momentum distribution of a quantum state of motion. \textbf{(b)} An initial state (green) is prepared in an optical tweezer. The state evolves over a time $t_e$. The tweezer is turned off from a depth of $V_f$, and the atom expands in free space (blue arrows) for a fixed flight time $t_f$. \textbf{(c)} The distribution measured after a given evolution time $t_e$ in the trap, $\tilde{p}(\theta)=\tilde{p}(\omega_x t_e)$, is a generalized quadrature measurement of the initial state. At specific $t_e$, this generalized quadrature can be equivalent to the momentum quadrature $p=\tilde{p}(0)$ or the position quadrature $x=(x_0/p_0)\tilde{p}(\pi/2)$.}
\end{figure*}

In this work, we demonstrate tomography of a single neutral atom prepared in non-classical motional states using time-of-flight imaging (Fig.~\ref{fig:Figure_1}). Time-of-flight samples a particle's momentum distribution (Fig.~\ref{fig:Figure_1}a), and has been used in optical tweezers to measure thermal single atoms~\cite{fuhrmanek2010imaging} and to probe spin correlations in few-fermion systems~\cite{bergschneider2019experimental}. Time-of-flight imaging and detection has also been used to great effect in neutral atomic gases and optical lattices~\cite{kurtsiefer1997measurement,greiner2002quantum,schellekens2005hanbury,bloch2008many-body,bucker2011twin}, and has enabled momentum distribution measurements of atomic ensembles prepared in squeezed, Fock, and superposition states~\cite{morinaga1999manipulation}, and tomography of the dynamics of a Bose-Einstein condensate~\cite{bucker2013vibrational}. In our work, using detection with single atom sensitivity, and measurements at multiple quadrature angles, we carry out full tomography and reveal negative valued Wigner functions with a single particle~\cite{romero-isart2011optically}.   

As illustrated in Fig.~\ref{fig:Figure_1}, to obtain a time-of-flight image we suddenly turn off the optical tweezer and allow the atom to fly in free space for a fixed time $t_f$.  We then average many such images to determine the momentum distribution at the time of release.  We extract arbitrary quadrature distributions by combining time-of-flight imaging with in-trap harmonic evolution for a time $t_e$ (Fig.~\ref{fig:Figure_1}b,c).  We start with a state $\psi_x(t_e=0)$ that we want to characterize, and measuring this initial state via time of flight gives the momentum quadrature $\tilde{p}(t_e=0)=p$.  If we allow the atom with mass $m$ to evolve in an ideal harmonic trap, the in-trap momentum after a time $t_e$ is the rotated quadrature $\tilde{p}(\theta) = p \cos \theta + (p_0 / x_0) x\sin \theta$ (Fig.~\ref{fig:Figure_1}c), where $\theta = \omega_x t_e$, and $x_0=\sqrt{\hbar/(2m\omega_x)}$ and $p_0=\sqrt{m\hbar\omega_x/2}$ are the characteristic length and momentum of the harmonic oscillator with angular frequency $\omega_x$. By varying the evolution time $t_e$ we can extract an ensemble of quadrature distributions $\tilde{p}(\theta)$ for $\theta \in [0,2\pi]$, analogous to what is done in optical homodyne tomography. The quadratures can be used to reconstruct the complete quantum state of the particle or equivalently the Wigner function $W(x,p)$.

In our experiments we test our protocol with multiple motional states, such as Fock states and displaced Fock states. To create the near-ground state of a single neutral $^{87}$Rb atom, we use Raman sideband cooling~\cite{kaufman2012cooling}. To then create non-classical motional states, we use versatile control of quantum tunneling in the optical tweezer~\cite{kaufman2015entangling}. This capability does not rely upon internal states and spin-motion coupling as in standard trapped ion settings~\cite{leibfried1996experimental,kienzler2015quantum}, and hence, can be extended to polarizable particles with no controlled internal degrees of freedom.  With large mass dielectric particles, while quantum tunneling is not experimentally feasible, it has been proposed that other non-harmonic potentials created by optical tweezers can enable quantum state synthesis~\cite{weiss2019quantum,ciampini2021experimental}.

Our experiments start by stochastically loading single $^{87} \text{Rb}$ atoms into optical tweezer traps using grey optical molasses and ascertaining the presence or absence of an atom through an initial in-trap fluorescence image. We then use optical molasses cooling followed by three-dimensional Raman sideband cooling in a trap of depth \SI{1.0}{\milli\kelvin} to prepare the atom close to the three-dimensional motional ground state $|n_x,n_y,n_z\rangle=|0,0,0\rangle$ (SM~\ref{appendix:tweezers})~\cite{kaufman2012cooling}. The trap is then adiabatically ramped down to a depth of \SI{0.33}{\micro\kelvin} where remaining thermal population is allowed to escape to further purify the initial state.  

\begin{figure}[tbh]\centering
\includegraphics[width=0.5\textwidth]{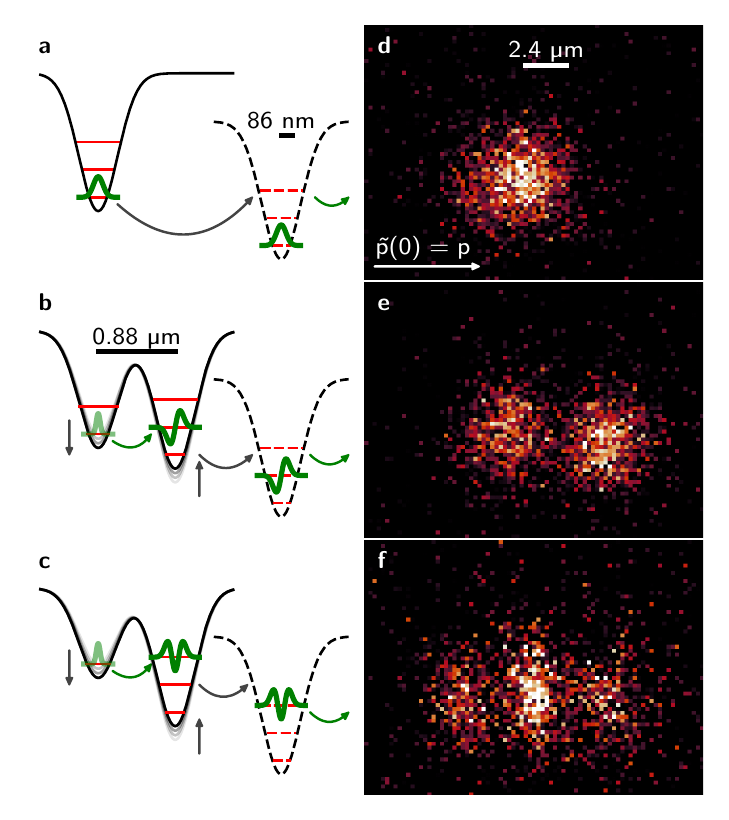}
\caption[Single-atom Fock state preparation and imaging.]{\textbf{Single-atom Fock state preparation and imaging. (a),(b),(c)} Illustration of motional state preparation of $n\in\{0,1,2\}$ states, respectively. \textbf{(d),(e),(f)} Time-of-flight momentum distributions at $t_e=0$ and for $t_f=0.5$ ms of $n_x=(0,1,2)$ states for which (64008, 48309, 58899) images were averaged, respectively.}
\label{fig:Figure_2}
\end{figure}

In the first set of experiments, we measure the momentum distribution of motional states at $t_e=0$ (Fig.~\ref{fig:Figure_2}).  After preparing the state of interest, we adiabatically ramp the trap to a final depth $V_f/k_B=\SI{2.4}{\micro\kelvin}$ and then abruptly turn off the trap (SM~\ref{appendix:tweezers}).  We then wait a flight time $t_f$ before applying resonant light for $\tau=\SI{10}{\micro\second}$ and collecting fluorescence through the high numerical aperture (NA) lens that creates the optical tweezers on an electron-multiplying CCD (EMCCD) camera. We repeat this procedure to realize multiple instances of single atom momentum measurements, and collect enough data to create an averaged momentum distribution that is observable above the camera noise. Experimental runs where an atom is not detected in the initial in-trap fluorescence image are used to characterize our imaging background, which is then subtracted from our captured momentum distribution (SM~\ref{appendix:imaging_characterization}).  

We first characterize the expansion dynamics of an atom prepared close to the ground state of the optical tweezer with angular trap frequencies $\omega_{x,y,z}$ (Fig.~\ref{fig:Figure_2}a,d). The initial ground-state root mean square (RMS) size $\sigma_{x,y}$ in position space is estimated as $x_0=$ \SI{86}{\nano\meter}, which is well below the resolution of our imaging system (Fig.~\ref{fig:Figure_1}a). At an expansion time of $t_f=\SI{0.5}{\milli\second}$, the atomic probability distribution has expanded to an RMS size of \SI{2.4(1)}{\micro\meter} in the atom plane, which is resolved by our imaging system (Fig.~\ref{fig:Figure_2}d).  By studying the expansion as a function of flight time $t_f$, we can ascertain that the expansion's kinetic energy observed in the radial directions of $(k_B\times\SI{0.256(16)}{\micro\kelvin})/2 $ is partly driven by the expected zero-point kinetic energy of the harmonic oscillator, $E_{\text{zp}}/2=\hbar\omega_{x,y}/4=(k_B\times \SI{0.188(1)}{\micro\kelvin})/2$  (SM~\ref{appendix:ballistic}). The difference in energy is due to the finite temperature of the atoms, which can also be seen in Raman spectroscopy (SM~\ref{appendix:raman}).

Starting with a ground-state atom, we create $n_x=1$ or $n_x=2$ motional Fock states using one-dimensional tunneling in a double well.  The tweezer is moved to \SI{0.88}{\micro\meter} from a second empty optical tweezer on the right, and with both tweezers near a depth of \SI{1.8}{\micro\kelvin} we bring the ground state of the left tweezer nearly energetically resonant with higher-$n_x$ states of the right tweezer. Then, with an adiabatic sweep of the relative tweezer depths, the atom is transferred into the target excited state of the right tweezer (Fig.~\ref{fig:Figure_2}b,c) (SM~\ref{appendix:tunneling})~\cite{kaufman2015entangling}.  The two tweezers are then slowly separated and the intensity of the left tweezer is ramped to zero, releasing any atom that did not successfully transfer to avoid polluting the final image (SM~\ref{appendix:image_analysis}).  We abruptly turn the remaining right tweezer off from \SI{2.4}{\micro\kelvin} and proceed with the same imaging procedure as for the $n_x=0$ state. The resulting $n_x=1$ and $n_x=2$ momentum distributions in Fig.~\ref{fig:Figure_2}e,f show characteristic fringing that is expected of the excited motional states. 

\begin{figure*}[tbh]\centering
\includegraphics[width=\textwidth]{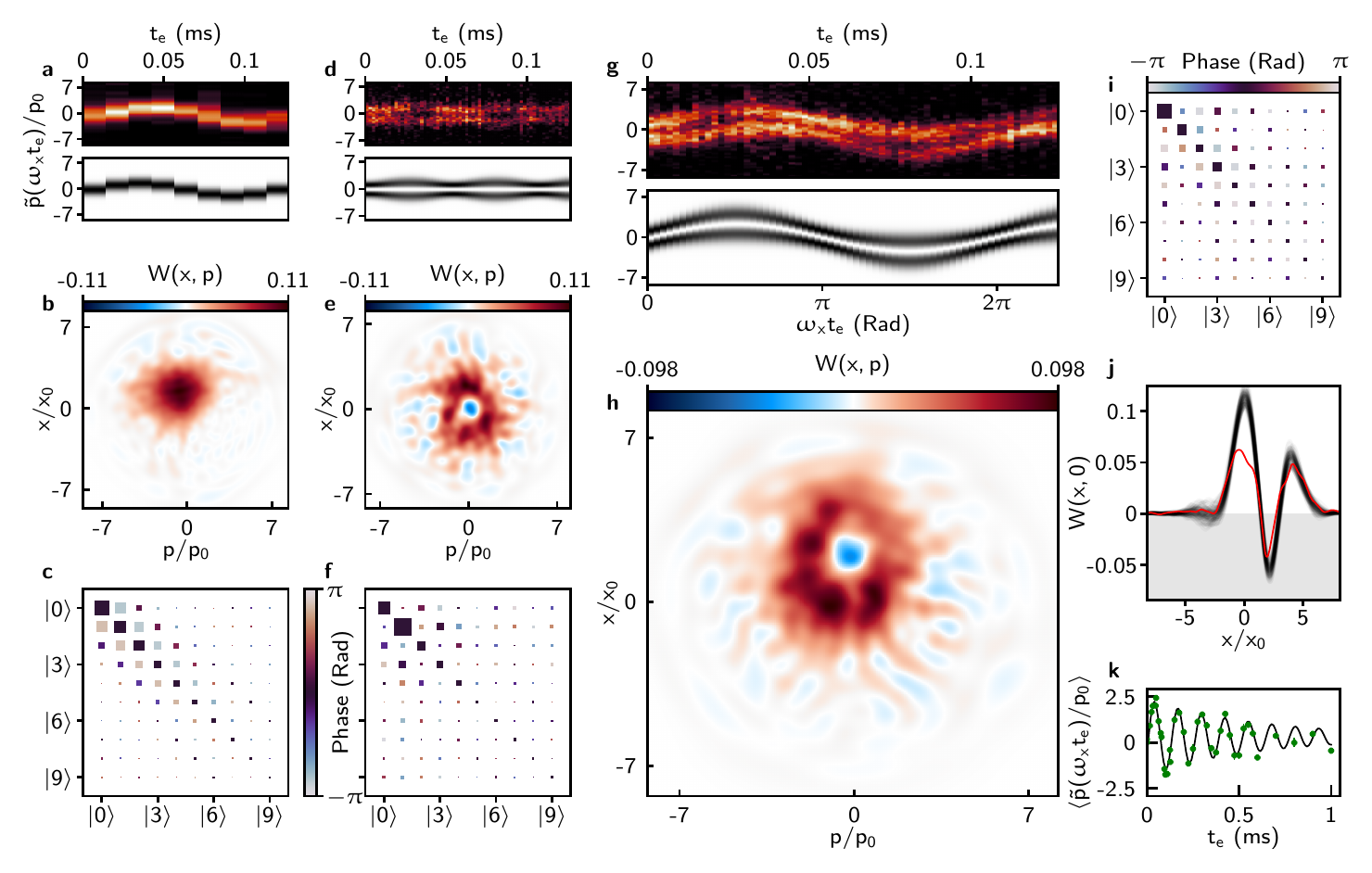}
\caption[Motional quantum state tomography via time-of-flight imaging and MLE]{\textbf{Motional quantum state tomography via time-of-flight imaging and maximum likelihood estimation}. Measured quadrature data, Wigner function, and density matrix Hinton plot for the displaced $n_x=n_y=0$ state \textbf{(a,b,c)}; $n_x=1$ state \textbf{(d,e,f)} and displaced $n_x=1$ state \textbf{(g,h,i)} both with a slight squeezing operator also applied. \textbf{Quadrature data:} Waterfall plots (a,d,g).  Each vertical slice corresponds to a raw quadrature distribution, such as in  Fig.~\ref{fig:Figure_2}, after deconvolving with the imaging PSF and integration along the vertical axis.  The measured data (upper waterfall plot) is compared to the expectation for ideal preparation and harmonic time evolution based upon our protocol (lower waterfall plot).  \textbf{Wigner functions:} Wigner functions (b,e,h) show classical positive values as red and non-classical negative values as blue.  We normalize the Wigner function such that the ideal negativity of a pure $n_x=1$ state is $-1/\pi$. \textbf{Density matrices:} Hinton plots (c,f,i) for density matrices as reconstructed via MLE. The area of each square is proportional to the magnitude of the corresponding element's complex value; the color of the square represents the element's phase. \textbf{(j)} $p=0$ slice of reconstructed Wigner functions from (h) (red). Equivalent slices as reconstructed through a bootstrapping method (black) that characterize the statistical uncertainty of our reconstruction algorithm. \textbf{(k)} Evolution of the measured center of the coherent state (green circles) and damped sinusoidal fit (black line), which is used to characterize the trap frequency and anharmonicity.}
\label{fig:Figure_3}
\end{figure*}

We now proceed to the full tomographic characterization of motional states. In these experiments, we study multiple quadrature distributions by waiting a variable amount of evolution time $t_e$ before releasing the atom and imaging the result.  We can visualize the quadrature data as time-sequence waterfall plots (Fig.~\ref{fig:Figure_3}a,d,g), which are derived from our raw distributions by deconvolving with the imaging point spread function (PSF) and integrating out the vertical axis (SM~\ref{appendix:image_analysis}).  We study states with a goal of testing the capacity of time-of-flight tomography to identify non-classicality and phase preservation, as well as subtle non-stationary features. First, we create a coherent state by starting with an $n_x=0$ state in the $V_f/k_B=\SI{2.4}{\micro\kelvin}$ trap and abruptly displacing the optical tweezer by \SI{180}{\nano\meter}. We find the state oscillates back and forth in the trap as expected (Fig.~\ref{fig:Figure_3}a).  Next, we produce an $n_x=1$ Fock state using the same protocol as the data presented in Fig.~\ref{fig:Figure_1}b.  In addition, after the tunneling and optical tweezer separation, the depth is suddenly doubled, resulting in application of a squeezing operation.  The atom is then released from a trap of $V_f/k_B=\SI{3.6}{\micro\kelvin}$. We observe that the state is mostly stationary, as expected, with the addition of a slight breathing from the squeezing induced by the depth jump (Fig.~\ref{fig:Figure_3}d). Lastly, we combine multiple techniques by starting with an $n_x=1$ state, applying the sudden doubling of the trap depth, and in addition displacing the optical tweezer by \SI{140}{\nano\meter}. As shown in Fig.~\ref{fig:Figure_3}g, we are able to observe the expected oscillation dynamics of the state in the trap.

To reconstruct the quantum state from the quadrature data, we choose to use maximum likelihood estimation (MLE)~\cite{leonhardt_measuring_1997} (SM~\ref{appendix:max_lik_estimation}).  An appropriately designed MLE algorithm takes the quadrature data and returns the density matrix that is most likely to reproduce this data~\cite{banaszek_maximum-likelihood_1999}. We implement an iterative MLE protocol based on the standard optical homodyne tomography literature~\cite{lvovsky2004iterative}, and from the density matrix, the Wigner function is directly recovered (Fig.~\ref{fig:Figure_3}). 

The results of applying the MLE algorithm to the quadrature data are presented in Fig.~\ref{fig:Figure_3}b,c,e,f,h,i. The coherent state displays significant off-diagonal coherences but, as expected, a positive Wigner function (Fig.~\ref{fig:Figure_3}b,c). The non-Gaussian state preparation associated with Fig.~\ref{fig:Figure_3}d,g results in the density matrices and Wigner functions with negative values displayed in Fig.~\ref{fig:Figure_3}e,h,f,i.  The value and statistical error on the density matrix and Wigner function negativity is estimated by a bootstrapping technique~\cite{lvovsky2004iterative}, in which we randomly sample predicted quadrature and noise distributions based on the MLE result and our camera noise characterization respectively. We then extract the density matrices of these data sets to create a statistical ensemble of density matrices and Wigner functions (SM~\ref{appendix:error_estimation}). The nearly-stationary $n_x=1$ state displays a dominant $n_x=1$ component and the Wigner minimum is found to be \SI{-0.060(6)} (Fig.~\ref{fig:Figure_3}e).  Adding a displacement demonstrates non-trivial off-diagonal coherences and a negative Wigner function value at the displaced center of \SI{-0.064(6)} (Fig.~\ref{fig:Figure_3}h,j). 

A full assessment of the reconstructed wavefunction must also consider systematic errors (SM~\ref{appendix:error_estimation}).  Trap anharmonicity, for example, will result in measured quadratures that do not simply follow from the rotated quadratures of the ideal protocol.  We determine the impact of anharmonicity by theoretically assessing the tomography protocol based on a model using measured trap parameters.  We estimate our trap anharmonicity by studying the coherent state oscillations of Fig.~\ref{fig:Figure_3}a over a longer time (Fig.~\ref{fig:Figure_3}k). The center of the Gaussian oscillates at \SI{7.84(5)}{\kilo\hertz}, and decays with a time constant of \SI{0.63(14)}{\milli\second}.  A model of the trap containing an anharmonic term is fitted to match the observed damping. For a displaced $n_x=1$ state in this model, we compare the reconstructed states obtained from MLE after evolution in a trap with and without our modelled anharmonic terms. We observe only a small infidelity of $<5\%$ between the resulting reconstructed states, and the Wigner function minimum for anharmonic evolution is smaller in magnitude by $<0.01$ compared to harmonic evolution, and remains negative (SM~\ref{appendix:Anharmonicity}).  In the future, the large dynamic range and control afforded by optical traps can be used to control the harmonicity.  Specifically, a shallow double-well or other anharmonic traps could be used for state creation, and the tomography could be carried out after ramping to a much deeper and less anharmonic trap. 

We have demonstrated quantum tomography of non-classical single atom motional states. By using time-of-flight imaging we have measured negative-valued Wigner functions with non-trivial phase-space structure. This work lays the foundation for tomography and characterization of massive levitated particles without exploitable spin structure~\cite{gonzalez2021levitodynamics,weiss2019quantum,vanner2013cooling}.  Further, time-of-flight imaging of single atoms will enable study of high-$n$ motional superposition states~\cite{mccormick2019quantum}, highly squeezed states, and interference~\cite{parazzoli2012observation} of complex delocalized states.

Acknowledgements:  We thank Tobias Thiele, Steven Pampel, and Ting-Wei Hsu for valuable insights and technical assistance, and Konrad Lehnert and Adam Kaufman for input on the manuscript.  We acknowledge funding from NSF Grant PHYS 1734006, ONR Grant N00014-17-1-2245 and Grant N00014-21-1-2594,  NSF QLCI Award OMA 2016244, and the U.S. Dept. of Energy, Office of Science, National Quantum Information Science Research Centers, Quantum Systems Accelerator, and the Baur-SPIE Endowed Professor at JILA. W.~J.~D acknowledges support from an NSF Graduate Fellowship.



\onecolumngrid
\vspace{30pt}


\begin{center}
\textbf{\large{Supplementary Materials}}
\end{center}

\begin{enumerate}[topsep=0pt,itemsep=0ex,partopsep=1ex,parsep=1ex,label=\Roman*.]
    \item Optical tweezers
    \begin{enumerate}[topsep=0pt,itemsep=-1ex,partopsep=1ex,parsep=1ex,label=\Alph*.]
        \item Tweezer generation and control via acousto-optic deflectors
        \item Tweezer loading
        \item Trap depth and frequency calibrations
        \item Excited Fock state preparation via tunneling
    \end{enumerate}
    \item Imaging methods
    \begin{enumerate}[topsep=0pt,itemsep=-1ex,partopsep=1ex,parsep=1ex,label=\Alph*.]
        \item Imaging setup
        \item In-tweezer RPGC imaging
        \item Time-of-flight imaging characterization
        \item Time-of-flight image analysis
    \end{enumerate}
    \item Single-atom temperature characterization
    \begin{enumerate}[topsep=0pt,itemsep=-1ex,partopsep=1ex,parsep=1ex,label=\Alph*.]
        \item Raman sideband spectra
        \item Ballistic expansion
    \end{enumerate}
    \item{Quantum state tomography and related characterizations}
    \begin{enumerate}[topsep=0pt,itemsep=-1ex,partopsep=1ex,parsep=1ex,label=\Alph*.]
        \item{Single-image Fock state momentum distribution analysis}
        \item {Maximum likelihood estimation algorithm}
    \end{enumerate}
    \item {Error estimation}
    \begin{enumerate}[topsep=0pt,itemsep=-1ex,partopsep=1ex,parsep=1ex,label=\Alph*.]
        \item{Estimating statistical error through bootstrapping}
        \item{Noise and imaging systamatic effects}
        \item{Trap anharmonicity}
    \end{enumerate}
\end{enumerate}

\section{Optical tweezers}
\label{appendix:tweezers}

\begin{figure*}[!ht]\centering
\includegraphics[width=\textwidth]{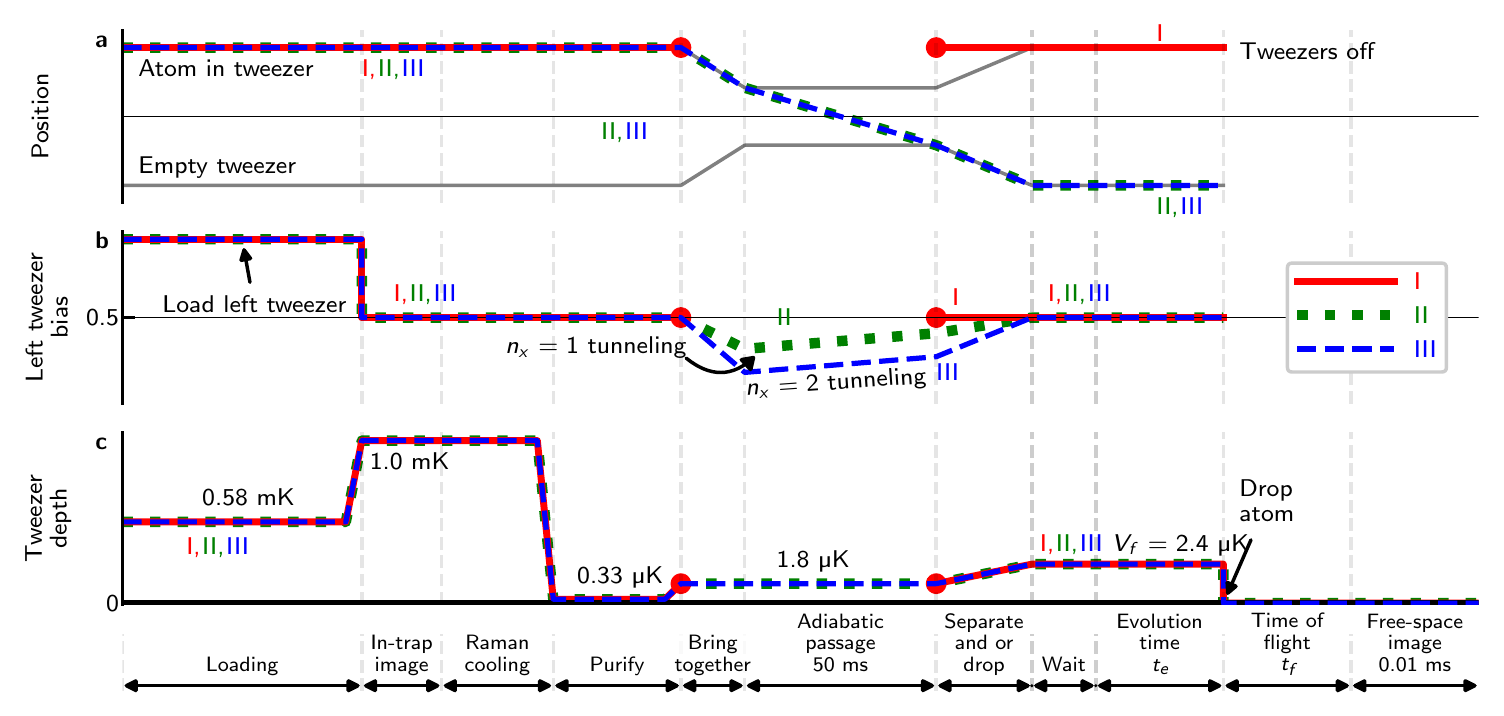}
\caption[Schematic Fock-state momentum distribution experiment timing diagrams]{\label{fig:timing1} \textbf{ Schematic timing diagram for Fock-state momentum distribution experiment (Fig.~\ref{fig:Figure_2})} Not-to-scale experiment diagrams, showing \textbf{(a)} the position of the atoms and tweezers, \textbf{(b)} the bias of the tweezers, and \textbf{(c)} the depth of the atom-holding tweezer. Stationary Fock-state momentum distributions in which we prepared $n_x=0,1,2$ states are labeled I, II, and III respectively. Round markers indicate adiabatic passage into an excited state is not used in experiment I that works with the $n_x=0$.
}\end{figure*}

\begin{figure*}[!ht]\centering
\includegraphics[width=\textwidth]{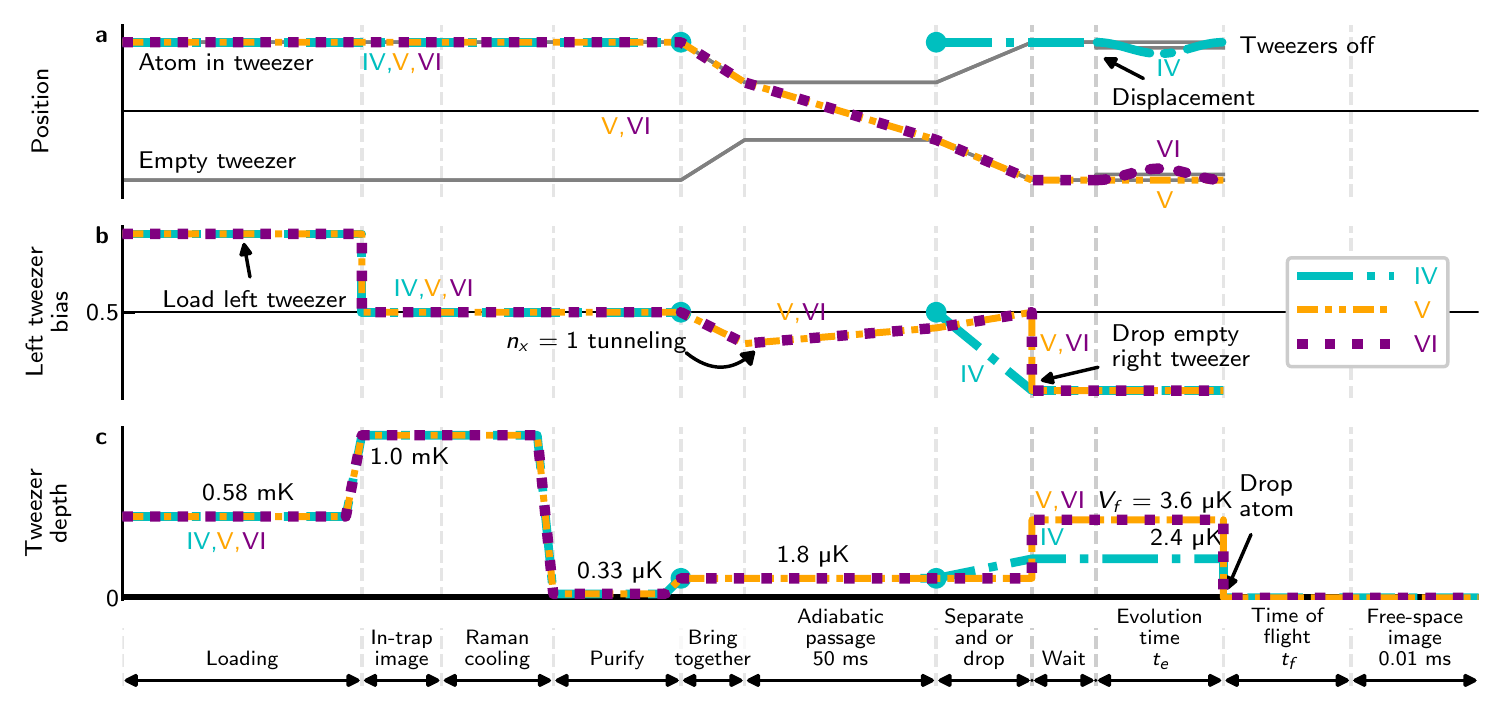}
\caption[Schematic tomography experiment timing diagrams]{\label{fig:timing2} \textbf{Schematic timing diagram for tomography experiment (Fig.~\ref{fig:Figure_3})} Not-to-scale experiment diagrams, showing \textbf{(a)} the position of the atoms and tweezers, \textbf{(b)} the bias of the tweezers, and \textbf{(c)} the depth of the atom-holding tweezer. Displaced coherent state data shown in Fig.~\ref{fig:Figure_3}a-c is labeled IV.  Non-displaced   $n_x=1$ data shown in Fig.~\ref{fig:Figure_3}d-f is labeled V. Displaced $n_x=1$ data shown in Fig.~\ref{fig:Figure_3}g-j is labeled VI. Round markers indicate adiabatic passage into an excited state is not used in experiment I that works with the $n_x=0$.
}\end{figure*}

\subsection{Tweezer generation and control via acousto-optic deflectors} 
\paragraph{Tweezer generation:} The optical tweezers are generated by sending light at a wavelength of \SI{850}{\nano\meter} through two orthogonal acousto-optic deflectors (AODs) that are driven simultaneously with multiple RF tones to create multiple deflections. In the case of this experiment, the horizontal AOD is used to generate the two deflections used for tunneling, in order to avoid complications arising from gravity. 

\paragraph{Depth and bias control:} The vertical AOD is used to direct extra laser power far away from the main tweezers that hold atoms, which allows us to reduce the depth of the main tweezers by many orders of magnitude even given the limited dynamic range of our intensity servo. The relative depths of the traps are modified by dynamically adjusting the amount of RF power in each tone driving the horizontal AOD.

\paragraph{Tweezer position and movement:} The tweezers are moved by changing the frequencies of the RF tones that drive the AODs. The speed of this movement is then limited by the size of the laser beam inside the AOD crystal and the time it takes the acoustic wavefront to cross this distance ($\sim\SI{100}{\nano\second}$). This is very fast compared to the tweezers' radial trap frequencies. However, there is \SI{10}{\micro\second} of electronic jitter in the time between the trigger to change the RF frequency and when the frequency jumps. This could be easily improved in future experiments, but is the likely explanation of small phase offsets noticeable in the center-of-mass oscillation data (Fig.~\ref{fig:COM_Oscillations}).

\subsection{Tweezer loading} During the initial loading stage, $V/k_B=\SI{0.58}{\milli\kelvin}$. The loading procedure is stochastic, and the $\Lambda$GM loading technique we use is capable of up to $\sim$90\% loading efficiency~\cite{brown2019gray}. However, because interleaving background images without atoms provides useful information, we use sub-optimal loading (50\% to 80\% efficiency) (Fig.~\ref{fig:timing1} and Fig.~\ref{fig:timing2}). 

\subsection{Trap depth and frequency calibrations} 
\paragraph{Trap depth calibration:} We calibrate our trap depth $V$ by measuring the light shift of the trap on the $D_2$ $^{87}$Rb $F=2\rightarrow F'=3$ transition at relatively high depths ($\sim\SI{0.1}{\milli\kelvin}$ to \SI{1}{\milli\kelvin}). We extrapolate this calibration to low depths where the shift is smaller than the linewidth of the transition and therefore difficult to measure directly.

\paragraph{Trap frequency calibration methods:} 
We can measure trap frequencies $\omega_{x,y,z}$ in deep traps through Raman sideband spectroscopy. We can then estimate the trap frequencies at smaller depths according to $\omega_{x,y,z}\propto\sqrt{V}$.

\begin{figure}[!ht]\centering
\includegraphics[width=0.55\textwidth]{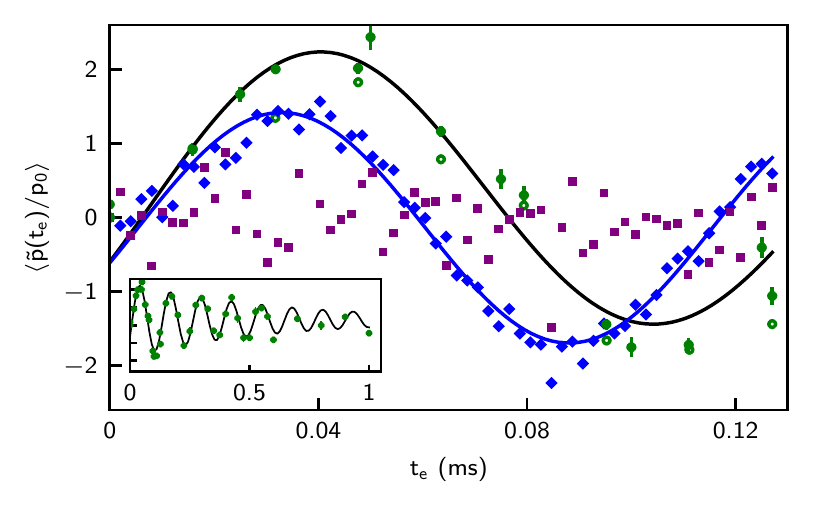}
\caption[Center of mass oscillations]{\label{fig:COM_Oscillations}
\textbf{Center of mass oscillations of data from Fig.~\ref{fig:Figure_3}.} 
Displaced $n_x=0$ centers of Gaussian fits (green circles), and for comparison $\langle\tilde{p}\rangle$ (open green circles) [in a \SI{2.4}{\micro\kelvin} trap].  Non-displaced $n_x=1$ $\langle\tilde{p}\rangle$ (purple squares), and displaced $n_x=1$ $\langle\tilde{p}\rangle$ (blue diamonds) [in a \SI{3.6}{\micro\kelvin} trap].   The $n_x=0$ ($n_x=1$) is fit with a decaying sinusoid to find an $n_x=0$ ($n_x=1$) oscillation frequency of \SI{7.84(5)}{\kilo\hertz} (\SI{9.05(11)}{\kilo\hertz}) (black and blue curves).  
Inset: Re-print of Fig.~\ref{fig:Figure_3}k for reference, which displays the same fitted centers of the $n_x=0$ data (green circles) and their fit (black curve) over a longer time.}
\end{figure}

Independently, displacing a state in a harmonic oscillator ideally causes the state's center-of-mass momentum to oscillate at the trap frequency. These oscillations can then be measured in order to independently determine the trap frequency at the small depths used for motional state preparation and tomography (Fig.~\ref{fig:COM_Oscillations}). The oscillations can be analyzed either by calculating $\langle\tilde{p}(t_e)\rangle$ from the deconvolved quadrature data (Fig.~\ref{fig:Figure_3}a,d,g), or they can be analyzed by fitting the quadrature data and tracking the location of the fit. Specifically for the displaced $n_x=0$ state, which is approximately Gaussian, the fits more reliably characterize the state as the fit is less susceptible to far-off-axis noise.

\paragraph{Comparing frequency calibrations:}
The $n_x=0$ state was released from a depth of $V_f=\SI{2.4}{\micro\kelvin}$. Extrapolating the Raman sideband spectroscopy trap frequency calibrations to this depth predicts $\omega_{x,y}/2\pi=\SI{6.8}{\kilo\hertz}$ and $\omega_{z}/2\pi=\SI{1.4}{\kilo\hertz}$. We measure via the center-of-mass oscillations of the displaced $n_x=0$ state \SI{7.84(5)}{\kilo\hertz} (Fig.~\ref{fig:COM_Oscillations}). 

The $n_x=1$ displaced and non-displaced states were released from a larger depth of \SI{3.6}{\micro\kelvin}, where we measure via the center-of-mass oscillations \SI{9.05(11)}{\kilo\hertz} for $n_x=1$ (Fig.~\ref{fig:COM_Oscillations}). Extrapolating the measured $n_x=0$ trap frequency to the larger depth predicts \SI{9.60(6)}{\kilo\hertz}. The difference between the value \SI{9.05(11)}{\kilo\hertz} and \SI{9.60(6)}{\kilo\hertz} is most likely due anharmonicity in these relatively shallow traps.  In Sec.~\ref{appendix:Anharmonicity}, we discuss a theoretical model for our trap that accounts for these observations.

\subsection{Excited Fock state preparation via tunneling}
\label{appendix:tunneling}

\paragraph{Tunneling parameters and calibrations:} In order to complete the adiabatic ground-to-excited state transfer (Fig.~\ref{fig:Figure_2}b,c), we must find and characterize the appropriate tunneling resonance and adiabatic transfer process. Tunneling is always done at a depth of $V=\SI{1.8}{\micro\kelvin}$ and with the tweezers spaced at a set Gaussian function spacing of \SI{0.88}{\micro\meter}. Assuming the tweezers are Gaussian, the spacing between the double-well minima is expected to be \SI{0.78}{\micro\meter} based upon this setting. For only the characterization of the resonances, we load atoms into both tweezers individually in order to measure both the intended transfer from ground to excited state and the unintended reverse transfer from excited state to ground state, which is a result of imperfect ground-state preparation.

We calibrate the relative tunneling depth ($\Delta V$) by comparing the width of a ground-to-ground tunneling resonance to its oscillation frequency. This is done at shallower depths where the ground-to-ground tunneling resonance is measurable. At the relatively deep depths used for ground-to-excited-state tunneling, the $n_x=0\rightarrow n_x=0$ tunnel coupling is too weak to be easily measured. We calculate where we expect it to be located based on the splitting between the $n_x=1$ and $n_x=2$ resonances and the assumption that the trap is harmonic, and set this location as $\Delta V=0$. 

\begin{figure*}[!ht]\centering
\includegraphics[width=\textwidth]{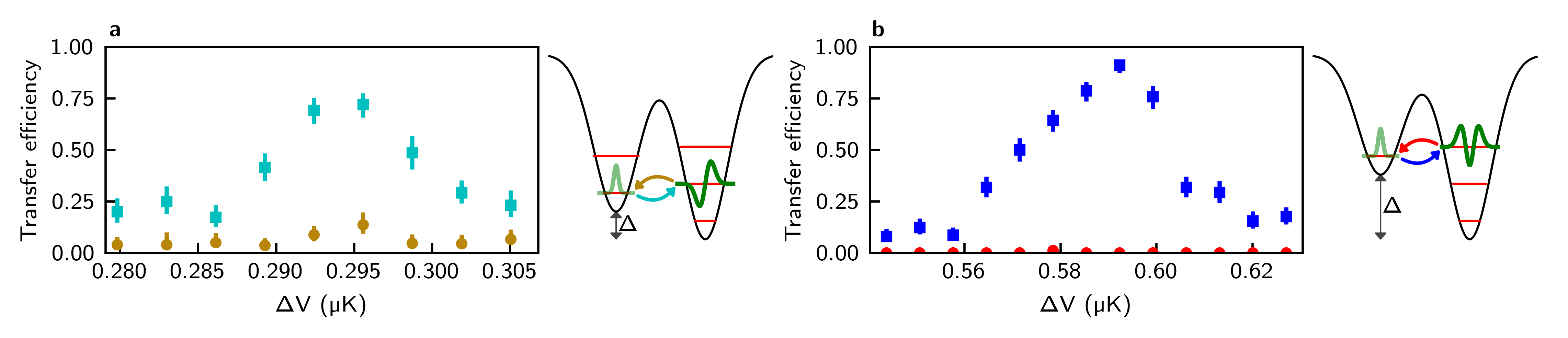}
\caption[Excited state tunneling resonances]{\label{fig:FSI_Tunneling_Both} \textbf{Excited state tunneling resonances.} \textbf{(a)} The $n_x=1$ tunneling resonance. Probability for an atom to tunnel from the ground state to $n_x=1$ (teal squares) and from $n_x=1$ to the ground state (gold circles). \textbf{(b)} The $n_x=2$ tunneling resonance. The probability for an atom to tunnel from the ground state to $n_x=2$ (blue squares) and from $n_x=2$ to the ground state (red circles). Due to the ground-state cooling there are nearly no atoms in the excited state of the second tweezer to transfer backwards.}
\end{figure*}

\paragraph{$n_x=0\rightarrow n_x=1$ characterization:} We characterize the $n_x=0\rightarrow n_x=1$ resonant tunneling transfer efficiency as a function of $\Delta V$. We fit the transfer efficiency with a Gaussian function and find that the resonance occurs at $\Delta V=\SI{294.5(2)}{\nano\kelvin}$ and has a RMS size of \SI{3.6(3)}{\nano\kelvin} (Fig.~\ref{fig:FSI_Tunneling_Both}a). Ramping the relative depths across a width of \SI{69.3}{\nano\kelvin} centered on this resonance over \SI{50}{\milli\second} achieves adiabatic rapid passage to the excited state at an efficiency of $92.4_{-2}^{+1.9}\%$. Meanwhile, we find transfer in the reverse direction, $n_x=1\rightarrow n_x=0$, to be $11_{-2}^{+3}\%$, reflecting that there is a small population in excited states of the tweezer, which is capable of transferring the reverse direction. This is consistent with our expectations from characterization of our single-atom temperatures via other methods where we estimate 90\% radial ground-state fraction (Sec.~\ref{appendix:raman}).

\paragraph{$n_x=0\rightarrow n_x=2$ characterization:} We similarly find an $n_x=0\rightarrow n_x=2$ tunneling resonance located at $\Delta V= \SI{589.0(11)}{\nano\kelvin}$ which has a RMS size of \SI{12.1(13)}{\nano\kelvin} (Fig.~\ref{fig:FSI_Tunneling_Both}b). 
Ramping the relative depths across a width of \SI{111}{\nano\kelvin} centered on this resonance over \SI{50}{\milli\second} achieves adiabatic rapid passage to the excited state at an efficiency of $86.0_{-3}^{+2}\%$. 
We find transfer in the reverse direction, $n_x=2\rightarrow n_x=0$, to be $3.2_{-1.1}^{+1.7}\%$, which is significantly smaller than the reverse transfer measured on the $n_x=0\rightarrow n_x=1$ resonance. This reflects that there is very little population in $n_x=2$ that is capable of transferring the wrong direction, as is expected after Raman sideband cooling.

\section{Imaging methods}\label{appendix:imaging_methods}

\subsection{Imaging setup}

The camera used is an Andor IXON-EM+ back-illuminated electron-multiplying CCD (EMCCD) camera, model number DU-897E-C00-\#BV-9GT. Camera settings for the experiment are listed in Table~\ref{tab:camera_settings_table}.  The imaging light at \SI{780}{\nano\meter} is collected through the same high-NA objective lens used to create our tweezer array, split from the optical tweezer light using a dichroic mirror, and focused using a \SI{1}{\meter} focal-length achromatic doublet. 

\begin{table}
\begin{center}
\begin{tabular}{ |l|l| }
\hline
 Parameter & Value \\ 
 \hline
 \hline
 Vertical shift speed & \SI{2}{MHz} \\
 Horizontal shift speed & \SI{1}{MHz} \\
 EM gain setting & $\times300$ \\
 Frame transfer mode & Off \\
 Pixel size & \SI{16}{\micro\meter}\\
 Pixel binning & 1x1 \\
 Camera temperature & \SI{-60}{\celsius} \\
 \hline
 Ideal imaging NA & 0.55\\
 Time-of-flight imaging time & \SI{10}{\micro\second} \\
 RPGC imaging time & \SI{5}{\milli\second} \\
\hline
\end{tabular}
\caption[Camera and imaging settings]{ \label{tab:camera_settings_table} \textbf{Camera and imaging settings.} (upper section) Camera settings directly programmed or reported by the manufacturer . Here, we report the vertical and horizontal shift speed as they are programmed. However, our camera is mounted sideways (Fig.~\ref{fig:Background_Image}).  (lower section) Imaging parameters.}
\end{center}
\end{table}

\subsection{In-tweezer RPGC imaging} 
\paragraph{RPGC imaging configuration:}
Red-detuned polarization gradient cooling (RPGC) and associated light scattering is used to determine if an atom is loaded at the start of the experiment. Time-of-flight images are post-selected based the presence of an atom in the RPGC image. Some time-of-flight images that record cosmic ray events are additionally removed in post-selection. Time-of-flight images where atoms are not loaded are used for background analysis. The RPGC cooling is in a balanced, $\sigma^+$-$\sigma^-$ configuration with zeroed background magnetic field. During the imaging, we alternate trap light and imaging light at a 2~MHz flashing rate~\cite{hutzler2017eliminating,lester2018measurement-based}. This eliminates light shifts and anti-trapping effects during the scattering of the imaging light to create more uniform images.

\begin{figure}[tbh]\centering
\includegraphics[width=0.5\textwidth]{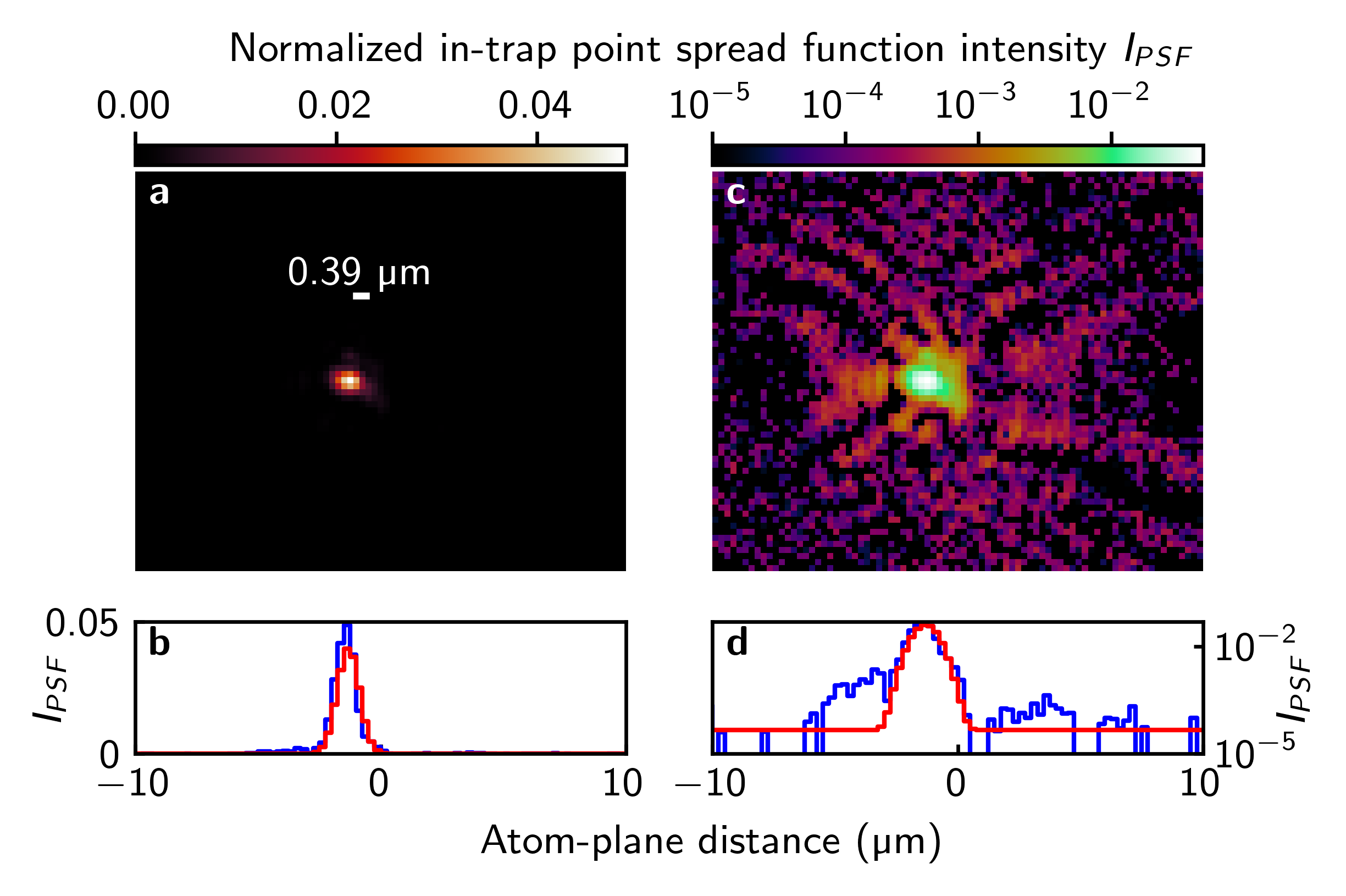}
\caption[The point spread function of our imaging system]{
\label{fig:psf}
\textbf{Imaging system point spread function}. Averaged image intensity (color bars) is normalized so that the integration of the averaged image is 1. The image shown is the average of 131,000 RPGC images with a single atom present. \textbf{(a)} The PSF on a linear color scale. 
\textbf{(b)} A slice of the PSF and a Gaussian fit to the PSF displayed on a linear $y$-axis. 
\textbf{(c)} The PSF displayed on a logarithmic color scale, emphasizing the observed long tail and structure due to aberrations. 
\textbf{(d)} A slice of the PSF and Gaussian fit to the PSF displayed on a logarithmic $y$-axis. }
\end{figure}

\paragraph{RPGC image point spread function:} 
One measure of the point spread function (PSF) of our imaging system is the average of in-trap RPGC images of single atoms (Sec.~\ref{appendix:image_analysis}). This averaged image represents the effect of the lens NA, aberrations, and the finite spatial extent of the atom in the trap during imaging. We note that this PSF may vary spatially within the field of view of the imaging system, but we utilize the central point as a representative value. The measured PSF (Fig.~\ref{fig:psf}) is roughly an astigmatic Gaussian with a long asymmetric tail. We subtract the averaged image's background and fit the result with a 2D Gaussian to extract effective RMS PSF sizes of \SI{0.445(2)}{\micro\meter} (horizontal) and \SI{0.328(2)}{\micro\meter} (vertical) (Table~\ref{tab:blurTable}).  This result is the PSF used for deconvolution in the tomography analysis. Considerations of systematic error based on this choice are discussed in Sec.~\ref{appendix:error_estimation}.

\subsection{Time-of-flight imaging characterization}\label{appendix:imaging_characterization}

\paragraph{Time-of-flight imaging configuration:}
The time-of-flight imaging is near-resonant and operates with an intensity $I\gg I_{\text{sat}}$. The light is applied in a power-balanced $\sigma_+,\sigma_-$ polarization configuration on the $D_2$ $f=2$ to $f'=3$ $^{87}$Rb transition with zero background magnetic field.

\paragraph{Collected photon number:}
\label{appendix:photon_number}
In a \SI{10}{\micro\second} image, as we nearly saturate the atomic transition, we expect to scatter 180 photons from a single atom. Based on the final quadrature distributions and characterizations of our camera system's count-to-photon conversion ratio at our EM gain setting (0.0124 photons per count), we estimate that we collect $\sim5-10$ photons, depending on which data set is analyzed, suggesting an overall collection efficiency of 3\%-6\%. 

\paragraph{Time-of-flight imaging blur considerations:}

In addition to the imaging resolution PSF discussed above, in time of flight there are additional blurring effects~\cite{fuhrmanek2010imaging,bergschneider2018spin}, and in this section we estimate these effects theoretically (Table~\ref{tab:blurTable}).

For $V_f/k_B=\SI{2.4}{\micro\kelvin}$, the confinement of the atom prior to release, combined with the measured atom temperature, we predict an initial atomic probability distribution RMS size of \SI{100}{\nano\meter}. During the $\tau=\SI{10}{\micro\second}$ imaging time, atoms move due to their initial velocity a RMS distance of \SI{57}{\nano\meter}. Atoms additionally undergo random-walk motion due to the scattering of the imaging photons. We estimate during $\tau$ the atom moves an additional RMS distance of \SI{0.83}{\nano\meter} ~\cite{fuhrmanek2010imaging}.

The released atomic wavefunction expands in 3D before being imaged onto the 2D image plane of our camera.  The extent of the atom probability distribution in the $z$ direction combined with a finite depth of focus is another potential blurring effect.  We apply geometric optics and the expected impact of diffraction to an impulse response of the form $\delta{x,y}\times\exp(-z^2/(2\sigma_z^2))$, where $\sigma_z$ here is the predicted $z$-size of the atomic wavefunction in our experiments after \SI{0.5}{\milli\second} of flight time. Diffraction alone would result in a PSF size of $\sigma_x=\SI{300}{\nano\meter}$, and our calculation indicates that the DOF increases this to $\sigma_x=\SI{320}{\nano\meter}$, which is \SI{110}{\nano\meter} as a value added in quadrature.

\begin{table}
\begin{center}
\begin{tabular}{ |l|l| }
\hline
 \multicolumn{2}{|l|}{\textbf{In-trap measurement}} \\
 \hline
 Measured PSF $\sigma_x$ & \SI{445(2)}{\nano\meter} \\ 
 Measured PSF $\sigma_y$ & \SI{328(2)}{\nano\meter} \\
 \hline
 \multicolumn{2}{|l|}{\textbf{Expectations}} \\
 \hline
 Expected diffraction limit $\sigma$ & \SI{300}{\nano\meter} \\
 In-trap thermal distribution $\sigma_T(0)$ & \SI{100}{\nano\meter} \\
 \hline
 \hline
 \multicolumn{2}{|l|}{\textbf{Atomic flight measurement}} \\ 
 \hline
  Measured PSF first \SI{10}{\micro\second} $\sigma_x$ & \SI{461(5)}{\nano\meter} \\
  Measured PSF first \SI{10}{\micro\second} $\sigma_y$ & \SI{366(4)}{\nano\meter} \\
 \hline
 \multicolumn{2}{|l|}{\textbf{Atomic flight calculations}} \\
 \hline
 DOF effect $\sigma_{DOF}$ & \SI{110}{\nano\meter} \\
 Thermal displacement during $\tau$, $\sigma_\tau$ & \SI{57}{\nano\meter} \\
 Random walk during $\tau$, $\sigma_{recoil}$  & \SI{0.83}{\nano\meter} \\ 
 \hline
\end{tabular}

\caption[Factors contributing to imaging blur]{\label{tab:blurTable}
\textbf{Factors contributing to imaging blur.}
The upper table reflects in-trap measurements and the expected diffraction limit calculated using the NA of the ideal tweezer lens. The lower table reflects blurring from effects unique to the time-of-flight imaging and that can be sensitive to the length of a single image $\tau$.
}
\end{center}
\end{table}

\paragraph{First \SI{10}{\micro\second} time-of-flight distribution measurement:}
A useful experimental comparison to the theoretical estimate of time-of-flight blur is the size of a time-of-flight distribution of a near ground state atom in its first \SI{10}{\micro\second} of flight, the smallest time-of-flight distribution possible without shortening the imaging time (Fig.~\ref{fig:FSI_Temperature}a). This distribution is affected by the photon-scattering random walk during a single image and some of the thermal displacement effects, although not DOF effects. Note, however, that it is a noisier PSF estimate than the in-trap RPGC based PSF estimate (Sec.~\ref{appendix:imaging_methods}) due to the significantly shorter single image time. We find this time-of-flight distribution has Gaussian RMS size of \SI{0.461(5)}{\micro\meter} and \SI{0.366(4)}{\micro\meter} in the horizontal and vertical directions respectively, which is only slightly larger than the in-trap measurement and is similarly astigmatic, indicating that the dominant source of blur is the imaging optics (Table~\ref{tab:blurTable}).  Error based on our ability to account all blurring effects is discussed in Sec.~\ref{appendix:error_estimation}.

\paragraph{Clock-induced charge noise:}
We are limited by so-called clock-induced-charge (CIC)~\cite{bergschneider2018spin}.
These are events in which a pixel on the camera gains a ``clock-induced" charge during the process by which photoelectrons are shifted into gain and readout registers. The CIC is amplified the same as any charge, making it indistinguishable from the photoelectrons we wish to measure. The number of such charges scales with the number of pixels in a single image. As such, there is a tradeoff between improved imaging resolution and camera noise. CIC is reduced by shifting electrons across the camera sensor and gain registers at the maximum rate. Even at the fastest shift speed of \SI{0.3}{\micro\second} per shift, Andor reports that CIC occurs at a rate of $1.8\times10^{-3}$ events/pixel.  Due to technical issues our Andor camera is only currently capable of operating at \SI{0.5}{\micro\second} per shift, which will increase CIC.

\begin{figure*}[tbh]\centering
\includegraphics[width=\textwidth]{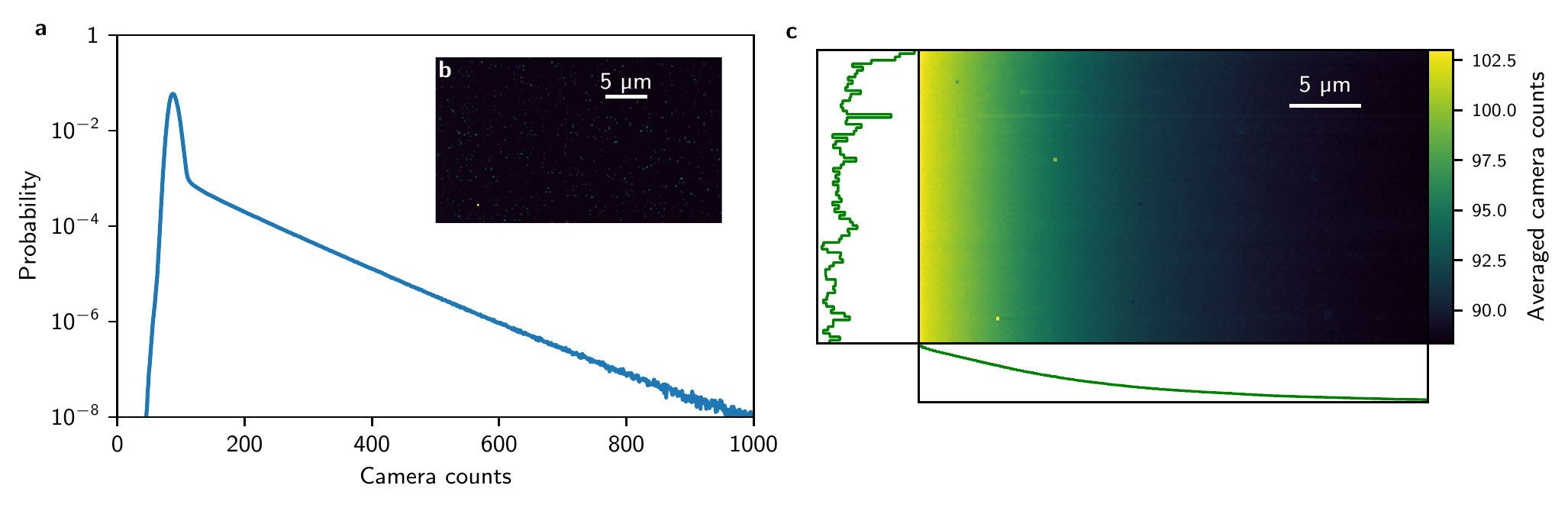}
\caption[Background analysis]{\label{fig:Background_Image} 
\textbf{Background and noise analysis.}
\textbf{(a)} The measured probability of observing a given count value in a single background image averaged over all columns in the images, which characterizes our camera noise.
\textbf{(b)} Inset: a representative, single, background image. High points in the image are primarily due to CIC.
\textbf{(c)} An averaged background, the average of many single background images. The CIC has mostly averaged out, and a non-negligible background profile has appeared, along with several hot spots. Vertical and horizontal integrations of the background are also displayed, emphasizing that most of the profile is in the horizontal direction (direction of Andor vertical shift, see Table~\ref{tab:camera_settings_table}), increasing toward the readout register on the left.}
\end{figure*}

\paragraph{Background characterization:}
An undesirable effect of averaging together many images is a sensitivity to background patterns on the camera, including ``hot" pixels that have consistently higher counts, a small amount of background light, patterns in the camera's sensitivity, large profiles in counts which are a result of the camera's cleaning and shifting processes, and CIC (Fig.~\ref{fig:Background_Image}). Most of the spatial variation in these noise sources is only horizontal, depending on the proximity of a column of pixels to the camera's readout sensor. Some of these sources vary over time. To combat these issues, backgrounds subtracted from our data are taken from the same experimental sequence as the data of interest, and are therefore temporally interleaved with the desired data. Most of these sources, such as the overall profile left by the cleaning process, are large compared to the averaged imaging signal, but very consistent. 

\paragraph{Gaussian camera noise characterization:} 
After subtracting the background from an averaged image but before further post-processing, there remains noise in the averaged image which is visible, for example, near the edges of the images in Fig.~\ref{fig:Figure_2}, where we expect negligible real signal. The noise on a single pixel approximately samples a Gaussian distribution and is primarily a result of camera CIC, so we refer to this as Gaussian camera noise.  Our analysis takes into account potential spatial dependence of the noise across the camera. 

We characterize this noise by first carefully measuring, from a very large data set of background images, the single-image count-probability distribution for every column of pixels (Fig.~\ref{fig:Background_Image}a).  If we analyze this noise probability distribution, we find an average CIC rate of $7.0 \times 10^{-2}$ events/pix, which is larger than ideal rates, largely due to our slower shift speed (Table~\ref{tab:noiseTable})~\cite{bergschneider2018spin}. With these column-wise-measured count-probability distributions, we can simulate an ideal uncorrelated background image by sampling from the measured single-image count-probability distributions. We can then create an averaged background by creating an arbitrary number $N$ of sampled single background images and averaging them together. We finally simulate many averaged backgrounds in order to reconstruct the probability distribution of a count value in an averaged background for every column of pixels. The resulting probability distribution approximately follows a Gaussian function with a standard deviation $\sigma_{\text{noise}}$, which decreases as $N$ increases. We simulate $\sigma_{\text{noise}}$ as a function of $N$, and fit the result with a function of the form $\sigma_{\text{noise}}=A/\sqrt{N}$. The resulting values of $A$ vary slightly by column, as expected, with a mean value averaged over all columns of $A=26.5$ counts and a standard deviation among the columns of 1.4. 

In practice, we find that the observed noise in our tomography data is larger than this sampling analysis would predict, as determined by analyzing the edges of background-subtracted quadrature distributions. Specifically, we observe RMS noise $(1.6,1.05,1.16)$ times larger for the $n_x=0$ displaced, $n_x=1$ non-displaced, $n_x=1$ displaced states, respectively, than the predicted noise from images with no atoms discussed above. The discrepancy is likely due to a combination of effects, such as differences in the background between different data sets and small correlations in the background data which average out in the larger background data set used for the above analysis. Non-trivial correlations exist in the background between data sets which are taken in close temporal proximity to each other, suggesting that some of the background signal may be due to fluctuating weak light signals, for example coming from light elsewhere in the apparatus. Some signal and noise may also come from atom-scattered photons as a result of our PSF having a long tail (Fig.~\ref{fig:psf}). To be consistent our observations, we always use the observed approximately Gaussian camera RMS noise from the tomography images when bootstrapping datasets.  We discuss error based on the amount of Gaussian camera noise in Sec.~\ref{appendix:error_estimation}.

\begin{table}
\begin{center}
\begin{tabular}{ |l|l| }
 \hline
 \multicolumn{2}{|l|}{\textbf{Specified camera noise}}\\
\hline
 CIC specified rate (ideal) & $1.8\times10^{-3} ~\text{events}/\text{pix}$ \\
 Readout noise & $< 1 e^{-}$ \\  
 Dark current & $1.5\times10^{-4} ~e^{-}/(\text{pix}\cdot \text{sec})$ \\
 \hline
 \hline
 \multicolumn{2}{|l|}{\textbf{Measured noise characteristics}}\\
\hline
 CIC measured rate  & $7.0\times10^{-2} ~\text{events}/\text{pix}$    \\
 Measured readout noise $\sigma$ & $5.4$ counts \\
 EM gain signal $\gamma$ & $73.1$ counts \\
 Count offset & $88.4$ counts \\
 \hline
\end{tabular}
\caption[Noise sources in our images.]{\label{tab:noiseTable}Noise sources in our images. Note that we expect our CIC rate to be larger than the specified value both due to our inability to run the camera at its fastest shift speed and because our measured value may be contaminated by a small amount of scattered light.
}
\label{table:noise}
\end{center}
\end{table}

\paragraph{Magnification:} \label{appendix:magnification}
We first measure the magnification of our imaging system using gravity in an atom-drop experiment. We prepare an atom in a tweezer, release it, and then take a picture a variable amount of time later to reconstruct the gravitationally assisted descent of the atom's center-of-mass position. We fit the signal with a Gaussian and plot the center as a function of flight time and fit to $y(t_f)=y_0+(1/2)a t_f^2$ to extract the acceleration $a$ of the magnified signal on the camera. The magnification is then $a/(\SI{9.8}{\meter/\second^2})$, which we find to be $\times64(1)$. This is the value used for calculations throughout the text.

Alternatively, we can measure the magnification based on measuring the $n_x=1$ Fock state. The distinctly non-Gaussian nature of the momentum distribution is a useful calibration because the fringe pattern is measurable in the image plane and independently predictable in the atom plane given $m$, $\omega_x$, and $t_f$.  In contrast to the Gaussian $n_x=n_y=0$ state, it is discernible from thermal population, which manifests as spatially-Gaussian-distributed signal with an RMS size dependent on the population's temperature. Specifically, we know how large the atom-plane $\sigma_x$ should be given the functional form of the $n_x=1$ state (Eq.~\ref{eq:phi_n0},\ref{eq:phi_n1},\ref{eq:phi_n2}), and we know from analyzing our momentum distributions (Sec.~\ref{appendix:MomentumDistributionImaging}) how large the image-plane $\sigma_x$ is. Taking the ratio of these two sizes gives us a magnification of $\times66$, which is consistent with our previous measurement.

\subsection{Time-of-flight image analysis} \label{appendix:image_analysis}

\paragraph{Imaging signal loss:}
In some cases, whether due to background collisions or thermal excitations, atoms that are imaged in the first RPGC image of the experiment are lost from the optical traps before the time-of-flight imaging procedure. In the case of optical homodyne tomography, where one typically measures the occupation of an optical mode, the loss of a photon registers as a measurement of the $n=0$ state. In our case, the loss of an atom does not result in the measurement of an atom in the $n=0$ state; rather, it simply results in a reduction of the signal to noise of the experiment, so no additional analysis is necessary to account for this.

\paragraph{Richardson-Lucy deconvolution:} 
The point spread function (PSF) of our imaging system is non-zero in size and astigmatic, resulting in blurring of the measured quadrature distribution. In order to compensate for these effects, we deconvolve our measured quadrature with an estimate of our PSF  (Sec.~\ref{appendix:imaging_methods}). While convolving two-dimensional imaging data is trivial, deconvolution is difficult numerically, being is very sensitive to any imaging noise.

We use the Richardson-Lucy deconvolution algorithm as implemented in the scikit-image python package. This iterative algorithm consists of iterating the relation

\begin{equation}
    D^{(k+1)} = D^{(k)} \times \Big( \frac{I}{D^{(k)}\circledast P } \circledast P^* \Big),
\end{equation}

\noindent where $D^{(k)}$ is the quadrature distribution after $k$ iterations of the algorithm, $I$ is the original distribution, $P$ is the PSF, $P^*$ is the PSF flipped along all axes, and $\circledast$ is the convolution operator. Multiplication and division of all terms is done element-wise. $D^{(0)}$ is initialized as an array uniformly filled with 1. This algorithm iteratively approaches the quadruatre distribution that, when convolved with our measured PSF, maximizes the likelihood of reproducing the measured quadruatre distribution, subject to Poissonian noise in our photon signal. This iterative method is similar to one that we use for density matrix reconstruction (Sec.~\ref{appendix:max_lik_estimation}).

In the equation above, a crucial intermediary step involves division, which in the presence of sufficient noise can produce issues related to the division by small numbers issues and floating point arithmetic. Therefore, the algorithm implements a filter step, whereby if an element in the term $I^{(k)}\circledast P$ is smaller than the filter value $\zeta_0$, the division result is floored to zero. While this retains the spirit of the algorithm, it also introduces added complexity into choosing the number of iterations to perform. As such, given the PSF there are two free parameters in the deconvolution algorithm: the number of iterations of the algorithm and the filter value $\zeta_0$. We discuss how we minimize error due to this in Sec.~\ref{appendix:error_estimation}.

\section{Single-atom temperature characterization}
\subsection{Raman sideband spectra} \label{appendix:raman}
A well-established method of characterizing the temperature of a single atom in a harmonic trap is Raman sideband spectroscopy that evaluates the imbalance between the red and blue Raman sidebands~\cite{kaufman2012cooling}. We do Raman sideband cooling at a trap depth of \SI{1}{\micro\kelvin} where we measure a radial trap frequency of \SI{139(4)}{\kilo\hertz} and an axial trap frequency of \SI{28(6)}{\kilo\hertz}. Our cooling is capable of producing 3D ground-state fractions greater than $90\%$. We estimate that over the course of several weeks of data taking required for this experiment $\bar{n}_x=0.10(4)$ in the radial dimension based on Raman sideband spectroscopy.  This corresponds to a temperature of $(k_B T/2)/(E_{\text{zp}}/2)=0.83(7)$, where $E_{\text{zp}}=\hbar\omega_x/2$ is the zero point energy. Our Raman sideband spectroscopy addresses both the $x$ and $y$-dimensions so we do not isolate the $x$-axis temperature. However, expansion momentum distributions indicate that there is no significant difference between radial axes (Sec.~\ref{appendix:ballistic}). The Raman spectroscopy temperature is consistent with our observation of $11_{-2}^{+3}$\% reverse transfer during g-e adiabatic tunneling transfer (Sec.~\ref{appendix:tunneling}). In the axial dimension, which we are relatively insensitive to, we measure $\bar{n}_z=0.08(10)$. 

\subsection{Ballistic expansion}\label{appendix:ballistic}
When the trap is shut off the atom's wavefunction will expand according to the standard ballistic expansion formula

\begin{equation}\label{eqn:ballistic_expansion}
\sigma_x(t) = \sqrt{\frac{2E_{\text{KE}}}{m}t_f^2 + \sigma_x(0)^2},
\end{equation}

\noindent for a given expansion kinetic energy  $E_{\text{KE}}$, which is the observable we directly measure via time-of-flight imaging. At high temperatures, $E_{\text{KE}}\rightarrow k_{B}T/2$. However, as $T\rightarrow 0$, the expansion energy diverges from the thermal energy as $E_{\text{KE}}$ is asymptotically dominated by the kinetic part of the harmonic-oscillator ground-state zero-point energy $E_{\text{zp}}/2=\hbar\omega_x/4$.

\begin{figure*}[!ht]\centering
\includegraphics[width=\textwidth]{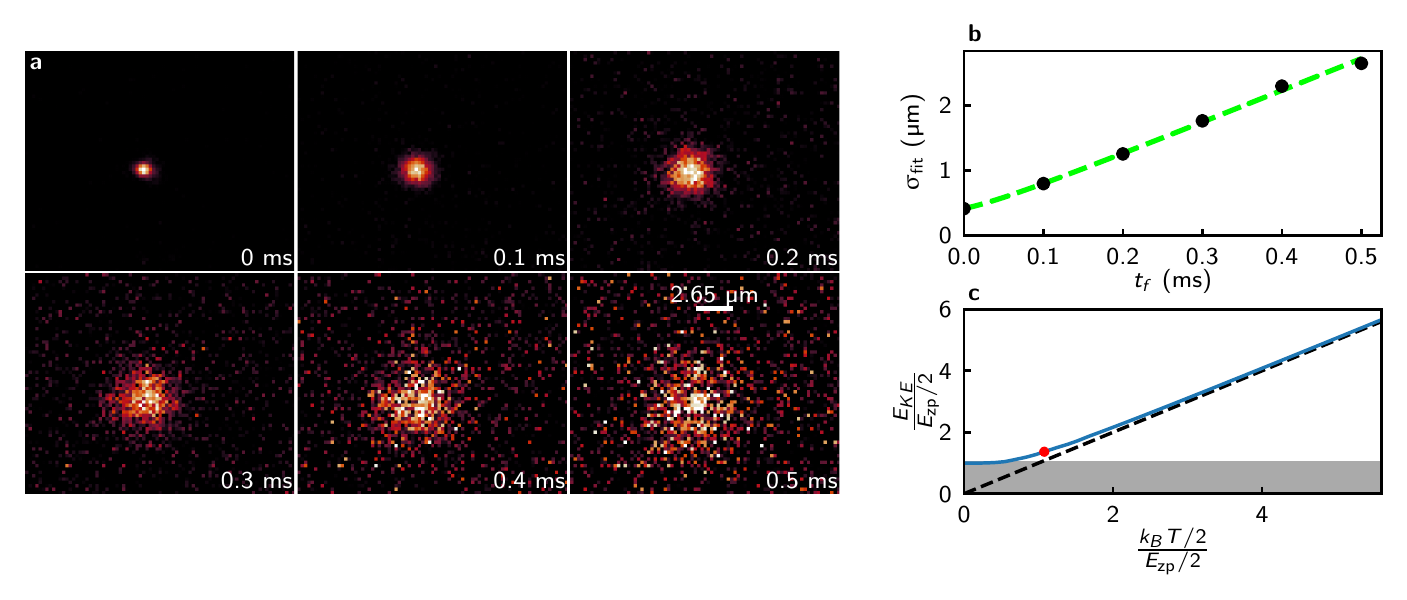}
\caption[Ground state expansion]{\label{fig:FSI_Temperature} 
\textbf{Ground state expansion.} 
\textbf{(a)} The averaged time-of-flight distributions used in the temperature measurement, labeled by the starting time of the \SI{10}{\micro\second} time-of-flight image. Color scales for different distributions are set individually. 
\textbf{(b)} The average RMS size $\sigma_{\text{fit}}=(\sigma_x+\sigma_y)/2$ of the 2D fits of the data in part (a) (black circles) and their ballistic expansion fit (green line) versus expansion time $t_f$. Error bars are too small to be visible.
\textbf{(c)} Relation between the measurable expansion kinetic energy $E_\textrm{KE}$ and the underlying thermal kinetic energy $k_B T / 2$ (blue line), a slope-1 reference (black line) and our measured data point (red point) which is significantly offset from the slope-1 reference, indicating that the zero-point energy is significant in this expansion.
}
\end{figure*}

To measure this expansion kinetic energy, we conduct a similar experiment as that of Fig.~\ref{fig:Figure_2}, but we vary the expansion time $t_f$ before imaging (Fig.~\ref{fig:FSI_Temperature}a). We fit the averaged momentum distributions with a 2D Gaussian function, and then fit the Gaussian RMS values as a function of expansion time with Eq.~\ref{eqn:ballistic_expansion} in order to extract an expansion energy from this data.

The experimentally measured results show a ballistic expansion kinetic energy of $E_{\text{KE}}=(k_B\times\SI{0.256(16)}{\micro\kelvin})/2$, which is close to the theoretical minimum at these trap parameters of $(k_B\times\SI{0.188(1)}{\micro\kelvin})/2$ (Fig.~\ref{fig:FSI_Temperature}). It makes no statistically significant difference whether we use $\sigma_x, \sigma_y, (\sigma_x+\sigma_y)/2$, or if we deconvolve the time-of-flight distributions with the in-trap PSF first with reasonable deconvolution parameters, therefore we quote the temperature obtained from analyzing the mean $\sigma$ of the non-deconvolved data $\sigma_{\text{fit}}=(\sigma_x+\sigma_y)/2$. This corresponds to a temperature of \SI{0.205(12)}{\micro\kelvin} (Fig.~\ref{fig:FSI_Temperature}c), which is significantly different than the ballistic expansion energy.  The temperature determined from ballistic expansion  corresponds to a normalized energy of $(k_B T/2)/(E_{\text{zp}}/2)=1.08^{+11}_{-10}$, which is similar to the value measured via Raman spectroscopy of $(k_B T/2)/(E_{\text{zp}}/2)=0.83(7)$~(Sec.~\ref{appendix:raman}). 
The slight discrepancy may suggest that the trap depth ramps that occurs between Raman cooling and the release of the atom for ballistic imaging are not perfectly adiabatic, or it may arise from fluctuations in the Raman cooling efficiency during the long experiments required to measure the atoms in free-space.

\section{Quantum state tomography and related characterizations}\label{appendix:tomography}
\subsection{Single Fock-state momentum distribution analysis}\label{appendix:MomentumDistributionImaging}

In general, the momentum space wavefunction is related to the position space wavefunction via a Fourier transform. The momentum space distribution of the $n_x=0, n_x=1,$ and $n_x=2$ states with ground-state RMS momentums $p_{0,x}=\sqrt{m\hbar\omega_x/2}$ and $p_{0,y}=\sqrt{m\hbar\omega_y/2}$ are
\begin{equation}
\label{eq:phi_n0}
|\phi_{0}(p_x, p_y)|^2 = \frac{1}{2\pi p_{0,x}p_{0,y}}\exp\Big(- \Big(\frac{p_x^2}{2p_{0,x}^2} + \frac{p_y^2}{2p_{0,y}^2}\Big)\Big),
\end{equation}
\begin{equation}
\label{eq:phi_n1}
|\phi_{1}(p_x, p_y)|^2 = \frac{1}{2\pi p_{0,x}^3 p_{0,y}} p_x^2\exp\Big(- \Big(\frac{p_x^2}{2p_{0,x}^2} + \frac{p_y^2}{2p_{0,y}^2}\Big)\Big),
\end{equation}
\begin{equation}
\label{eq:phi_n2}
|\phi_{2}(p_x, p_y)|^2 = \frac{1}{4\pi p_{0,x}p_{0,y}} \Big(\frac{p_x^2}{p_{0,x}^2}-1\Big)^2\exp\Big(- \Big(\frac{p_x^2}{2p_{0,x}^2} + \frac{p_y^2}{2p_{0,y}^2}\Big)\Big).
\end{equation}

As a baseline estimate of the stationary state, we can fit the raw momentum distributions from Fig.~\ref{fig:Figure_2}d-f with a variable-weighted sum of the above momentum-space distributions convolved with our imaging PSF to estimate the populations in the different states we prepare. We scale the RMS $x$ and $y$ sizes of the above $n_x=0$ ($n_x=1$ and $n_x=2$) states based on our displacement-based trap-frequency measurement of the $n_x=0$ ($n_x=1$) states, $t_f$, and the independently measured imaging magnification of $\times 64(1)$. It is important to fix the RMS size for characterization of the ground-state, as a ground-state Gaussian of one trap frequency can be easily confused with a thermal state with a smaller trap frequency. The excited state momentum distributions, which do not suffer from this issue, can be used as independent calibrations of the magnification system (Sec.~\ref{appendix:image_analysis}), but for this analysis we fix the size of these functions as well. 
This analysis is a useful preliminary diagnostic of our state preparation and imaging system. The $n_x=0$ state was measured to have $n_x=(0,1,2)$ populations of $(0.93(3), 0.07(2), 0.00(2))$ which is consistent with our expectations based on Raman sideband spectroscopy (Sec.~\ref{appendix:raman}), free-space expansion (Sec.~\ref{appendix:ballistic}), and excited state preparation (Sec.~\ref{appendix:tunneling}). 
The $n_x=1$ state was measured to have diagonal populations of $(0.260(10), 0.651(14), 0.089(15))$, and the $n_x=2$ state had diagonal populations of $(0.395(14), 0.125(18), 0.480(16))$. Deviations of these excited state results from the ideals of $(0,1,0)$ and $(0,0,1)$ likely result from a combination of imperfect state preparation (Sec.~\ref{appendix:tunneling}) and blurring effects in the imaging which are not accounted for by the PSF deconvolution (Sec.~\ref{appendix:imaging_methods}). Most such blurring effects that reduce the fringe contrast naturally manifest as population in lower states, so these numbers can be reasonably understood as lower-bound estimates on our Fock state preparation procedure.  These considerations are treated in more detail in the context of the full tomography in Sec.~\ref{appendix:error_estimation}.

\subsection{Maximum likelihood estimation algorithm}\label{appendix:max_lik_estimation}

We now turn to analysis of the full tomographic quadrature data of  Fig.~\ref{fig:Figure_3}.

\paragraph{Quadrature data:}
The data in Fig.~\ref{fig:Figure_3}a,d,g for displaced $n_x=0$, non-displaced $n_x=1$, and displaced $n_x=1$ states consist of data sets of 9, 64, and 64 quadrature measurements respectively. Each quadrature measurement is the average of a variable number of pictures. The median numbers of pictures averaged for these individual quadrature measurements were 10852, 6191, and 11320 pictures for the data in Fig.~\ref{fig:Figure_3}a,d,g. The non-displaced $n_x=1$ data was taken with notably fewer pictures than the other data sets.

We note that $N$ quadrature measurements are needed to distinguish phase oscillations in the quadrature distributions differing in angular frequency by $N\omega_x$, assuming the quadrature measurements are equally spaced by phase angle $2\pi/N$. Hence, $N$ such measurements are needed to reconstruct a density matrix occupying Fock states up to $n_x = N/2$~\cite{leonhardt_measuring_1997}. However, even in the case where we have occupation of states with $n_x \geq N/2$, we expect to accurately capture the density matrix up to the $N/2$ off-diagonal. Given the expected form of the density matrix, considering independent measures of the atom temperature and the size of the trap displacements, we estimate the number of quadrature measurements used is sufficient for capturing all non-trivial elements of the density matrix within error bounds.

\paragraph{MLE algorithm:}
Maximum likelihood estimation (MLE) is a common statistical technique for estimating the free parameters of a model based on how likely they are to reproduce a given data set. In the case of quantum tomography, we can take the quantum state as a model for producing measured observables, where the free parameters are typically the complex-valued matrix elements of the density matrix in some basis. MLE will then return the density matrix most likely to result in the observed data. This approach is attractive as we may a priori impose physical assumptions about our state, as opposed to other methods like the inverse Radon transform that may predict nonphysical states~\cite{lvovsky2004iterative}.

To determine the density matrix, we utilize an iterative MLE algorithm based off of the discussion in Ref.~\cite{lvovsky2004iterative}, which we briefly review here. Suppose we have a set of projective measurement outcomes, $\{(\tilde{p}_j,\theta_j)\}$, with shorthand $\tilde{p}=\tilde{p}(\theta)$, labelled by index $j$ and occurring with frequency $f_j$ in our data set. Here, $\tilde{p}_j$ is the result of a measurement of $\tilde{p}({\theta_j}) = p\cos\theta_j + m\omega_x x\sin\theta_j$, corresponding to the quadrature distribution at phase angle $\theta_j$. For any given pair $(\tilde{p},\theta)$, we may form the projection operator $\Pi\left(\tilde{p},\theta\right) = \ket{\tilde{p},\theta}\bra{\tilde{p},\theta}$ so that the probability of obtaining such a measurement in the state $\rho$ is given by 
\begin{align}
    P\left(\tilde{p},\theta\right) = \mathrm{Tr}\left(\rho\,\Pi\left(\tilde{p},\theta\right)\right).
\end{align}
The problem of MLE then corresponds to finding the quantum state $\rho$ that maximizes the likelihood of obtaining the data set $\{(\tilde{p}_j,\theta_j)\}$,
\begin{equation}
    \mathcal{L}\left(\rho\right) = \prod_j P\left(\tilde{p}_j,\theta_j\right)^{f_j}, \label{eq:likelihood}
\end{equation}
while also subjecting the resulting state to various physical constraints. Namely, $\rho$ must be a trace-normalized, Hermitian, positive semi-definite matrix in a convenient, physically-motivated basis of our choosing.

An iterative algorithm to maximize Eq.~\eqref{eq:likelihood} is motivated by the observation that, for the state $\rho_0$ maximizing $\mathcal{L}$, we must have $f_j \propto P\left(\tilde{p}_j,\theta_j\right)$ in the large sample limit. Introducing the operator
\begin{equation}
    R\left(\hat{\rho}\right) = \sum_j \frac{f_j}{P\left(\tilde{p}_j,\theta_j\right)}\Pi\left(\tilde{p}_j,\theta_j\right),\label{eq:R}
\end{equation}
and noting that $\sum_j \Pi\left(\tilde{p}_j,\theta_j\right) \propto I$ for identify matrix $I$, this translates to the condition
\begin{equation}
    \rho_0 \propto R\left(\rho_0\right)\rho_0R\left(\rho_0\right).
\end{equation}
We may then leverage this to form the basis of an iterative algorithm,
\begin{equation}
    \rho^{(k+1)} \propto R\left(\rho^{(k)}\right)\rho^{(k)}R\left(\rho^{(k)}\right),
\end{equation}
where we start with the initial trial state $\rho^{(0)} \propto J$, where $J$ is a matrix of ones in our truncated Fock basis, which has a nonzero probability for each possible outcome. We iterate this procedure until the error, as defined via $T\left(\rho^{(k)},\rho^{(k+1)}\right)$ for trace distance $T\left(\rho,\rho^\prime\right) = \frac{1}{2}\text{Tr}(\sqrt{(\rho - \rho^\prime)^2})$ is less than $10^{-4}$ or 500 iterations have been completed.

In practice, it is necessary to choose an appropriate reduced basis in which to restrict the set of possible states. For our purposes, we restrict ourselves to motional Fock states with occupation $n \leq n_{\textrm{max}}$ for a maximum occupation $n_{\textrm{max}}=25$. The projection operators can then be defined via
\begin{equation}
    \begin{split}
    \Pi_{mn}\left(\tilde{p},\theta\right) &= \bra{m}\Pi\left(\tilde{p},\theta\right)\ket{n} \\
    &= \langle m | \tilde{p},\theta\rangle \langle \tilde{p} ,\theta| n \rangle
    \end{split}
\end{equation}
with the standard harmonic oscillator matrix element

\begin{equation}
\langle n | \tilde{p},\theta \rangle = ie^{-in\theta} \Big(\frac{1}{2\pi p_0^2}\Big)^{1/4} \frac{H_n\left(\tilde{p}/p_0\sqrt{2}\right)}{\sqrt{2^n n!}}e^{-(\tilde{p}/p_0)^2/4}
\end{equation}
for Hermite polynomials $H_n$.

After reconstructing the density matrix, the state's Wigner function may be obtained via
\begin{equation}
    W(x,p) = \frac{1}{\pi}\sum_{m,n=0}^{n_{\textrm{max}}} \rho_{mn}(-1)^k e^{-i\phi(m-n)}e^{-|\alpha|^2/2}\sqrt{\frac{k!}{(k+\Delta)!}}|\alpha|^\Delta L_k^{\Delta}(|\alpha|^2)
\end{equation}
for $\alpha = x/x_0+i p/p_0$, $\phi = \arg(\alpha)$, $k = \min(m,n)$, and $\Delta = |m-n|$, where $L_k^\Delta$ are the generalized Laguerre polynomials.

\section{Error estimation}
\label{appendix:error_estimation}

\subsection{Estimating statistical error through bootstrapping} 

To present the Wigner functions distributions in Fig.~\ref{fig:Figure_3} we apply MLE directly to the observed quadrature data. When this is done, we obtain Wigner function maximum negativities of -0.052 (-0.043) for the non-displaced (displaced) $n_x=1$ states discussed in Fig.~\ref{fig:Figure_3}. However, this calculation does not naturally lend itself to estimation of the error on these values or on the density matrices. 

The calculation of confidence intervals for the results of the MLE algorithm is a non-trivial task. We take the bootstrapping approach outlined in Ref.~\cite{lvovsky2004iterative}. We begin by calculating the expected quadrature distributions as a function of $\theta$ and $\tilde{p}$ for the non-bootstrapped MLE density matrix, which is obtained from our measured quadrature data. For every value of $\theta$, we then simulate projective measurements of $\tilde{p}$ by sampling from these expected quadrature distributions, and use this to generate an ensemble of bootstrapped images.

Separately, based on our Gaussian camera noise analysis, we reconstruct noise that is added to each bootstrapped quadrature distribution. Specifically, we use Gaussian-distributed noise with a standard deviation consistent with what we observe in the tomography quadrature data (Sec.~\ref{appendix:imaging_characterization}). Note, for the highly-averaged statistics of our experiment, the quadrature distributions are well-sampled, and the projection noise is small compared to the Gaussian camera noise.

At this point, we have an ensemble of bootstrapped quadrature distributions, which we can run through our deconvolution algorithm and MLE, as done for our raw experimental data. This yields an ensemble of density matrices, whose variation reflects the Gaussian camera noise in the quadrature data.

\begin{figure*}[!ht]\centering 
\includegraphics[width=0.95\textwidth]{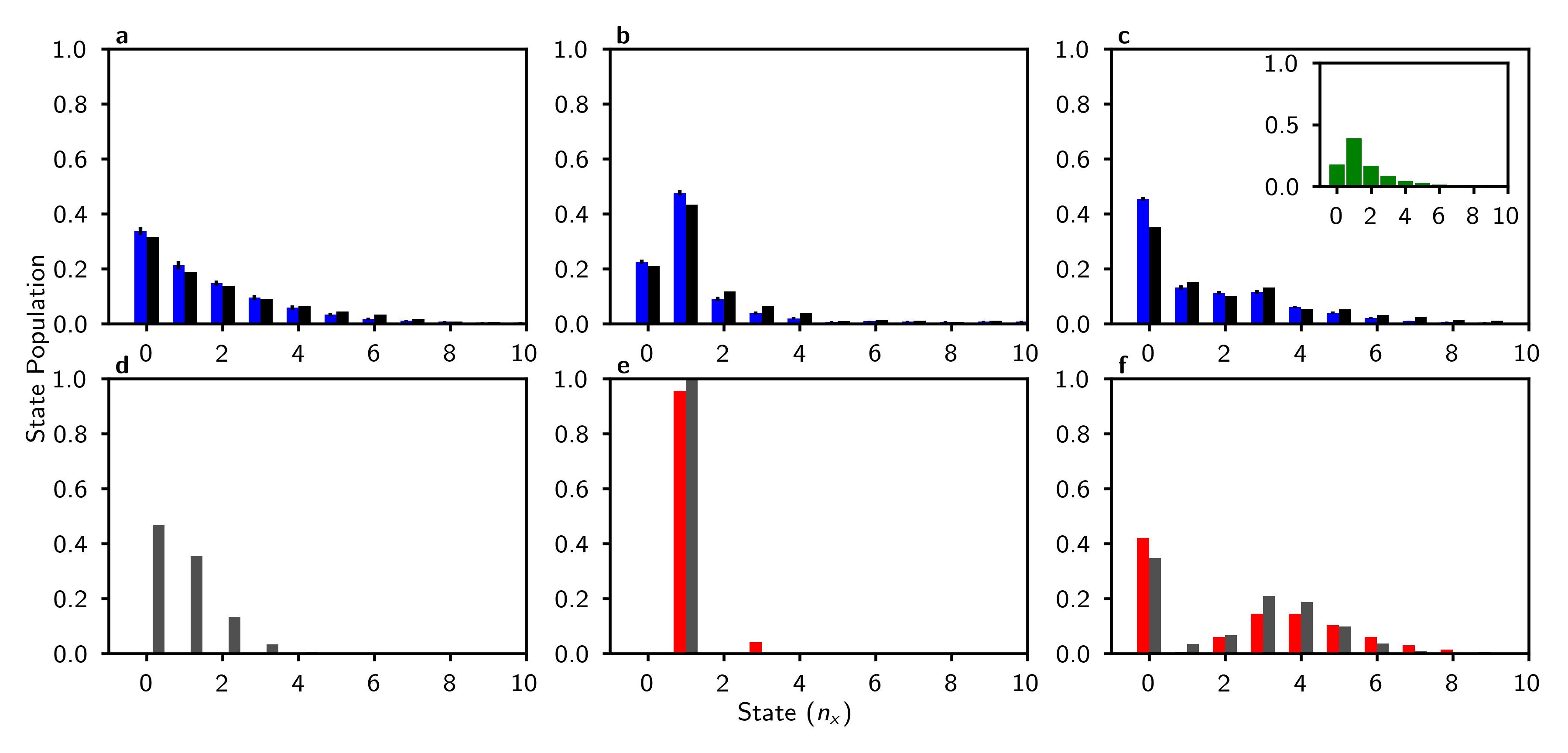}
\caption[State Populations Visualization]{
\textbf{State populations comparison via diagonal populations} (top row) Diagonal populations of harmonic oscillator states measured in our tomography as calculated through the error bootstrapping (blue bars) or the direct MLE result (black bars) for \textbf{(a)} the displaced $n_x=0$ state, \textbf{(b)} the non-displaced $n_x=1$ state, and \textbf{(c)} the displaced $n_x=1$ state. \textbf{(c)} inset: populations of the experimentally displaced $n_x=1$ data set after being displaced back to the origin.
(bottom row) For comparison, we display the populations of theoretically ideal states, which do not include experimental imperfections in state preparation, with (red bars) and without (grey bars) the applied squeezing operator for \textbf{(d)} the displaced $n_x=0$ state, \textbf{(f)} the non-displaced $n_x=1$ state, and \textbf{(e)} the displaced $n_x=1$ state.
}
\label{fig:population_bar_plots}
\end{figure*}

The diagonal density matrix elements derived from the direct MLE and the bootstrapped ensemble are displayed in Fig.~\ref{fig:population_bar_plots}a,b,c. While off-diagonal components are also important for the state inference, this picture provides a simple method to compare different techniques and expectations.  From this comparison we can see the bootstrapped ensemble results in slightly less population in higher $n_x$ states. From these density matrices we calculate the Wigner negativities, as presented in the main text. The minimum value of the Wigner function for the nearly-stationary $n_x=1$ state is found to be \SI{-0.060(6)} and the displaced $n_x=1$ state of \SI{-0.064(6)}. Recall, for the direct MLE analysis we obtain Wigner function maximum negativities of -0.052 (-0.043) for the non-displaced (displaced) $n_x=1$ states discussed in Fig.~\ref{fig:Figure_3}.  While all analyses clearly point to negative Wigner values, we note the difference between the non-bootstrapped and bootstrapped analyses is likely indicative of subtle effects such as non-ideal noise and the non-invertibility of our numerical deconvolution routine (Sec.~\ref{appendix:imaging_characterization}). 

Lastly, we note that our reconstructed Wigner function for the displaced and non-displaced $n_x=1$ states are qualitatively similar despite having independent noise patterns (Fig.~\ref{fig:Figure_3}d,g), which indicates the robustness of the MLE reconstruction protocol.

\subsection{Noise and imaging systematic effects}

\begin{figure}[!ht]\centering 
\includegraphics[width=\textwidth]{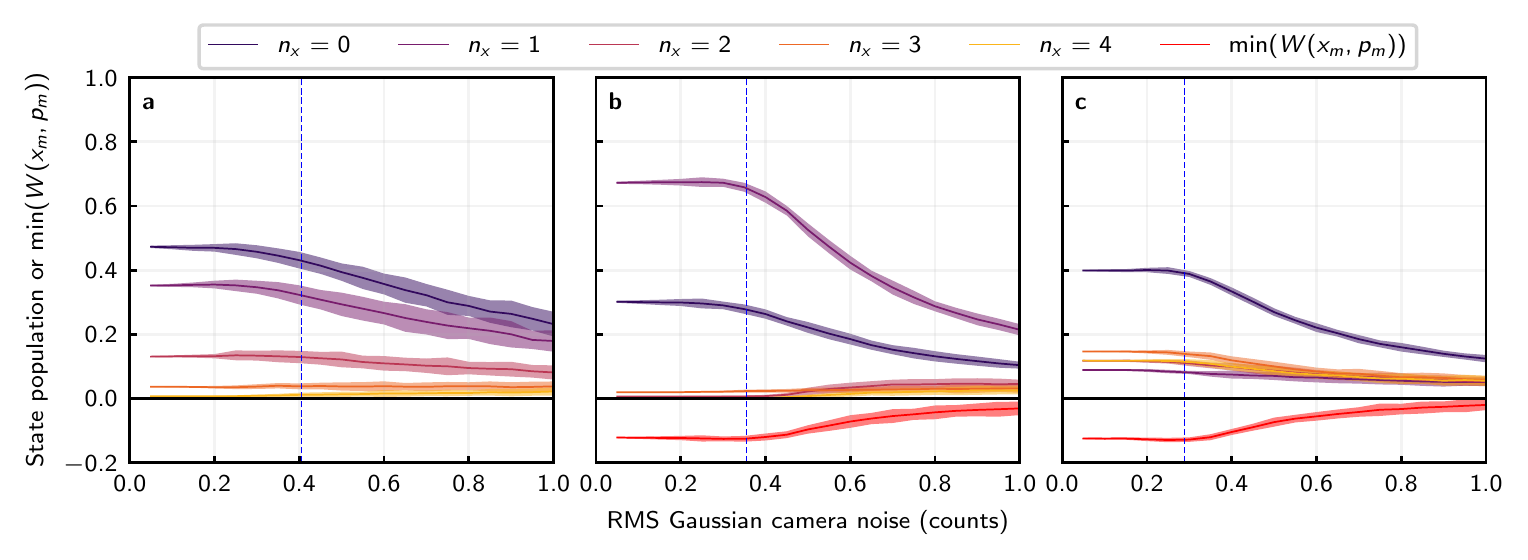}
\caption[Simulating the Effects of Gaussian Noise on Maximum Likelihood Estimation]{
\textbf{Simulating the effects of Gaussian noise on maximum likelihood estimation}
\textbf{(a)} Simulation of noise on a numerically displaced $n_x=0$ state with undisplaced diagonal $n_x=(0,1,2)$ populations of  $(0.93(3), 0.07(2), 0.00(2))$ and no off-diagonal populations, the values suggested by least squares fitting of the highly-averaged momentum distributions in Fig.~\ref{fig:Figure_2}, for demonstration. Shown are the mean values (solid lines) of populations $n_x=0$ through $n_x=4$, averaged over 50 simulations at each RMS noise value. 95\% of the single-simulation observed values fall within the color bands to demonstrate variation of these parameters between simulations. 
\textbf{(b)} Simulation of noise on an undisplaced but numerically squeezed $n_x=1$ state with undisplaced and unsqueezed $n_x=(0,1,2)$ populations of  $(0.260(10),0.651(14),0.089(15))$ picked similarly from the least squares fitting routine for demonstration. Additionally shown are the mean Wigner values at the location of the true minimum $(x_m,p_m)$ (red line), and 95\% of single-simulation Wigner function values at this location lie within the red color band. 
\textbf{(c)} Similar to (b) with the same initial populations, but for a displaced $n_x=1$ state.  A key reference point is the measured Gaussian camera noise, as measured in our real quadrature data in Fig.~\ref{fig:Figure_3} for each state (blue dotted lines).
}
\label{fig:supp_noise_sim}
\end{figure}

\paragraph{Systematic error due to variable Gaussian noise:}
We can simulate the systematic effect of varying amounts of Gaussian noise on our MLE state reconstruction by again using our bootstrapping procedure.  For this simulation, we construct mixed states with only diagonal elements based on the least squares fits of the heavily averaged momentum distributions in Fig.~\ref{fig:Figure_2}, as this data has the least relative noise. By adding variable amounts of Gaussian noise to this initial state and conducting MLE on the result, we find that increasing amounts of Gaussian noise biases the reconstruction to overestimate the population in higher-$n$ states, and underestimates the true negativity of the Wigner function (Fig.~\ref{fig:supp_noise_sim}).

\paragraph{Point-spread function and blurring effects:}
As discussed in Sec.~\ref{appendix:imaging_methods}, in order to accurately reconstruct our state, we must take into account aberrations and blurring effects in the time-of-flight images. Ideally, one could measure and characterize all such effects to reconstruct the true point-spread function, which can then be used to deconvolve the measured quadrature distribution. Because characterizing all blur is challenging for certain types of effects, we use a conservative value in our analysis, namely the in-trap size measured with RPGC, which is our smallest PSF estimate (Sec.~\ref{appendix:imaging_methods}). A less conservative estimate is the larger time-of-flight distribution of the atoms immediately after being released.

We verify that using larger PSF functions for deconvolution in the analysis of displaced and non-displaced $n_x=1$ Fock state data (Fig.~\ref{fig:Figure_3}d-j), up to a factor of 1.5 larger in both dimensions, results in uniformly more negative Wigner functions. For example, using the first-\SI{10}{\micro\second} time-of-flight distribution (Table~\ref{tab:blurTable}), which also has slightly less astigmatism, has a negligible impact on our Wigner negativity --- the reconstructed non-displaced $n_x=1$ Wigner function would have a minimum value from the non-bootstrapped MLE analysis of -0.055 (instead of -0.052).

\paragraph{Other deconvolution parameters:}
In general, we expect the optimal deconvolution parameters to depend on the nature of the state being reproduced and the amount of noise present. To choose suitable values for our analysis, we perform bootstrapping similar to that done for statistical error estimation, except that we allow the deconvolution parameters in the bootstrapped ensemble to vary from the direct MLE reconstruction parameters.  We then compare the resulting bootstrapped density matrix to the direct MLE result to find optimal parameters that reproduce the non-bootstrapped density matrix with the highest possible fidelity, where the fidelity is defined as
\begin{equation}
    \mathcal{F}=\Bigg(\text{Tr}\bigg(\sqrt{\sqrt{\rho_1}\rho_2\sqrt{\rho_1}}\bigg)\Bigg)^2.
    \label{eq:fidelity}
\end{equation}
Let $\zeta_0$ and $\zeta_b$ denote the deconvolution filter values for our direct MLE and boostrapped density matrices, respectively. We choose $\zeta_0$ such that the boostrapped density matrix with filter value $\zeta_b$ yields the highest fidelity when $\zeta_b = \zeta_0$.  For the displaced $n_x=1$ data set, we find that a filter value of $\zeta_b=0.69$ counts and between 2 and 10 iterations of the deconvolution algorithm reliably reproduces the density matrix for a wide range of $\zeta_0$. We thus choose to use this value, $\zeta_0=0.69$, in our main analysis. To minimize additional artifacting, we also apply only 2 iterations of the deconvolution algorithm, for which our bootstrapping routine reproduces the non-bootstrapped density matrix with a fidelity of 0.934(2).

We apply a similar analysis to find that the highest-fidelity parameters for the displaced $n_x=0$ analysis also corresponds to a filter value of $\zeta_0=0.69$, and that the best for the non-displaced $n_x=1$ is $\zeta_0=1.03$, both similarly with 2 iterations of the algorithm. The non-displaced $n_x=1$ requires slightly larger filtering to compensate for having slightly increased noise due fewer points in this data set.

\paragraph{Noise and imaging analysis summary:} 

In conclusion, noise and imaging blur processes can influence the MLE algorithm when carrying out tomography using imperfect camera imaging.  We have shown through a survey of systematic effects that their impact is either small or they result in our state reconstruction being a conservative underestimate of parameters such as the magnitude of the Wigner function negativity and state preparation fidelity. 

\subsection{Trap anharmonicity}
\label{appendix:Anharmonicity}

\begin{figure}[!ht]\centering 
\includegraphics[width=0.5\linewidth]{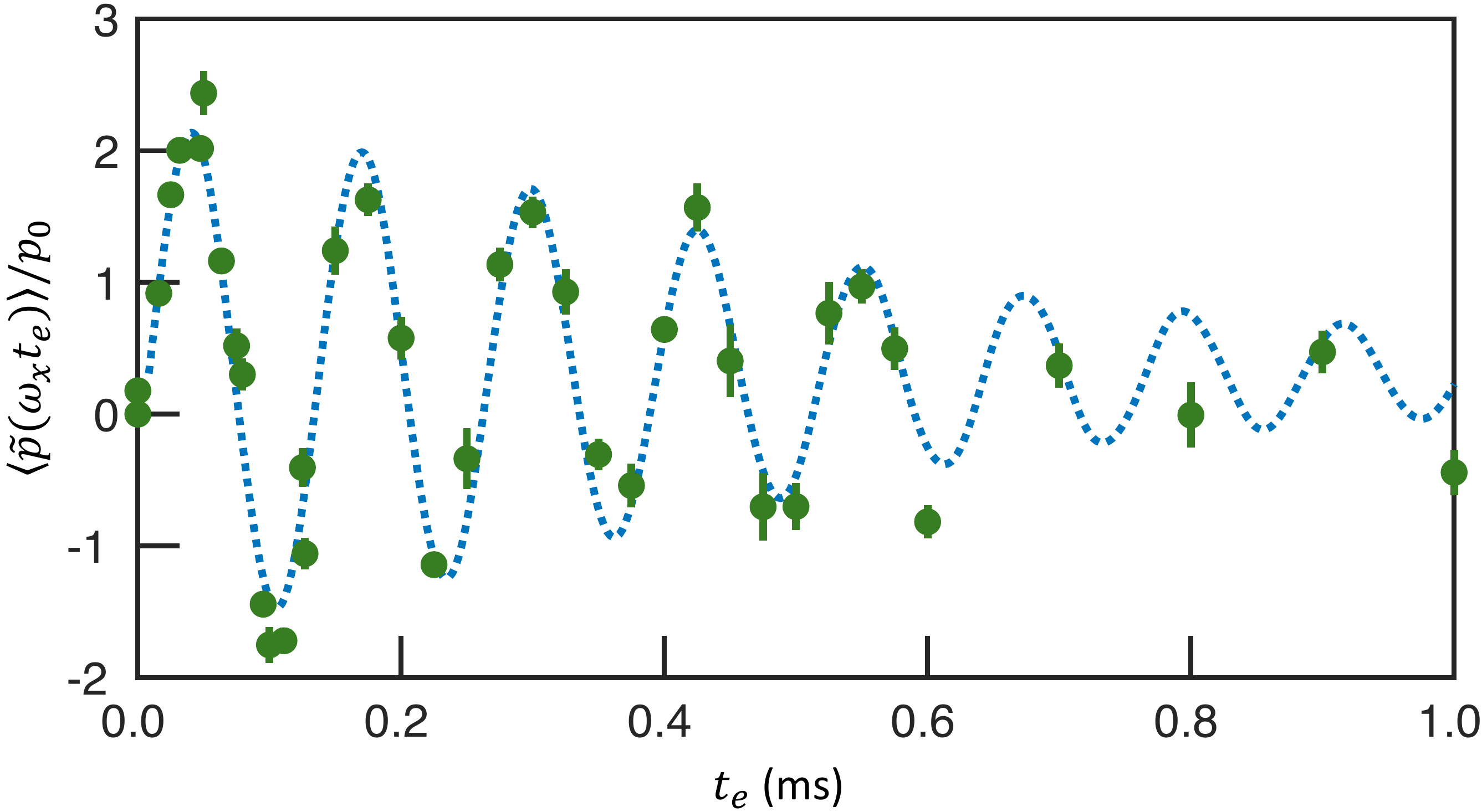}
\caption[Fits of an anharmonic oscillator to center of mass oscillations]{
Best-fit dynamics for $\langle \tilde{p}(\omega_x t_e)\rangle/p_0$ generated by Eq.~\eqref{eq:H_quartic} for an initial displaced ground state (blue, dotted), compared to experimental results from Fig.~\ref{fig:COM_Oscillations} for an initial displaced $n_x=0$ state (green). 
}
\label{fig:supp_COM}
\end{figure}

\paragraph{Characterizing trap anharmonicity}
To analyze the effect of trap anharmonicity on state reconstruction, we consider the dynamics of a single atom in a one-dimensional anharmonic oscillator, which serves as a phenomenological model for the experimental optical tweezer potential. Specifically, we consider dynamics generated by the Hamiltonian
\begin{equation}
\begin{split}
    H =& \frac{p^2}{2m} + \frac{m\omega_x^2x^2}{2} + \Lambda x^4 \\=& \hbar\omega_x\left[(p x_0/\hbar)^2 + \frac{1}{4}(x/x_0)^2 + \lambda (x/x_0)^4\right] ,
\end{split}
\label{eq:H_quartic}
\end{equation}
which describes an atom of mass $m$ in a harmonic trap with frequency $\omega_x$, plus an additional quartic term of characteristic strength $\Lambda$. In the second line, we have rewritten this potential in terms of the harmonic length $x_0 = \sqrt{\hbar/(2m\omega_x)}$, and defined the dimensionless parameter $\lambda = \Lambda x_0^4/(\hbar \omega_x)$ as the relative strength of the anharmonic correction.

We can obtain estimates of $\omega_x$ and $\lambda$ by comparing the predictions of our phenomenological model with experimental observations. Specifically, we compare experimental data for the long-time center-of-mass dynamics of an initially-displaced $n_x=0$ state (see Fig.~\ref{fig:Figure_2}k) to the average momentum $\langle \tilde{p}(\omega_x t_e) \rangle$ obtained from Eq.~\eqref{eq:H_quartic} for an initially-displaced $n_x=0$ state. In the case that $\lambda < 0$, we add a minimal sextic term $\hbar\omega_x(2\lambda^2/3)(x/x_0)^6$ that ensures our potential remains stable and retains only a single local minimum, and thus induces only local deformations of our ground state; this term plays a negligible role in the resulting dynamics for the considered parameters.

Treating $\omega_x$, $\lambda$, and the initial state displacement $x_i$ as free parameters, we perform a least-squares fit of $\langle \tilde{p}(\omega_x t_e)\rangle $ to the experimental center-of-mass oscillations. We obtain best-fit values of $\lambda = -0.0037(4)$, $\omega_x = 2\pi \times \SI{8.50(5)}{kHz}$ (corresponding to $x_0 = \SI{83}{nm}$) and $x_i = \SI{166(5)}{nm}$. As shown in Fig.~\ref{fig:supp_COM}, we observe that the fitted model robustly captures the observed damping. As a crude estimate of the relative significance of the harmonic and quartic potential terms in the Hamiltonian when considering the dynamics of this displaced state, we can compare their expected values at $t_e = 0$. By approximating moments by their harmonic values, we find $4\vert\lambda\vert\langle (x/x_0)^4\rangle /\langle (x/x_0)^2\rangle \approx 0.13$. While this indicates that the harmonic term remains dominant for displacements of this size, the anharmonic term is not so small as to be easily discounted without further analysis. Separately, we note the the best-fit estimate for the trap frequency in this model ($\SI{8.50(5)}{kHz}$, with $x_0 = \SI{83}{nm}$) is comparable to the observed oscillation frequency in the experiment, $\SI{7.84(5)}{kHz}$ with $x_0 = \SI{86}{nm}$. The deviation between these values is consistent with the negative $\lambda$ in our model, which generally leads to a slower observed oscillation frequency for a displaced $n_x=0$ state than what is expected for the corresponding ideal harmonic oscillator.

\paragraph{Maximum likelihood estimation in an anharmonic potential}
We now examine how anharmonicity affects state reconstruction, utilizing the Hamiltonian in Eq.~\eqref{eq:H_quartic} as a model for the tweezer potential. Following the protocol in the main text, we spatially displace an initial state $\rho_i$ by a fixed amount $x_i$, which in the Schr{\"o}dinger picture can be described by acting the displacement operator $D(x_i) = e^{-ip x_i/\hbar}$ on the initial state. We then evolve for a time $t_e$, described by the unitary operator $U\left(t\right) = e^{-it_eH/\hbar}$. The resulting dynamical state is expressed as,
\begin{align}
    \rho\left(t_e\right) = U\left(t_e\right)D(x_i)\rho_iD^\dagger(x_i)U^\dagger\left(t_e\right)\label{eq:rho_tau},
\end{align}
from which we extract the time-evolved momentum distribution $P\left(p,t_e\right) =  \mathrm{Tr}\left(\rho\left(t_e\right)\ket{p}\bra{p}\right)$ for momentum eigenstate $\ket{p}$. This distribution is used as the input for the iterative MLE algorithm (Sec.~\ref{appendix:max_lik_estimation}).

To make connections with the experiment, we would like to analyze the effect of an anharmonic trap on reconstruction of a displaced $n_x=1$ state. Because the prepared state contains contributions from other Fock states, we assume the initial state $\rho_i$ can be well modelled by an incoherent mixture of the low-energy eigenstates of the Hamiltonian,
\begin{align}
    \rho_i = P_0 \ket{0}\bra{0} + P_1\ket{1}\bra{1} + P_2 \ket{2}\bra{2} . \label{eq:rho0}
\end{align}
Here, $P_{n_x}$ denotes the probability of staring in the eigenstate $\ket{n_x}$; we denote this state by the triplet $\mathbf{P} = (P_0,P_1,P_2)$. By choosing $P_{n_x}$ that generally resembles the makeup of the state in the experiment we are able to analyze the effects of initial mixed states in both the dynamics and ensuing state reconstruction, and demonstrate robustness of MLE tomography over a range of possible initial states. 

As the state reconstruction data for the displaced $n_x=1$ state, featured in Fig.~\ref{fig:Figure_3}g-k, is obtained using a deeper trap depth and smaller displacement than that used in our initial analysis of the anharmonic oscillator model Eq.~\eqref{eq:H_quartic} for the $n_x=0$ center-of-mass oscillations, we modify the previously obtained best-fit parameters by an appropriate scaling. Assuming that $H$ is linear in the trap depth $V$, we have that $\lambda \sim 1/\sqrt{V}$ and $\omega_x \sim \sqrt{V}$. Hence we can rescale based on the relative ratios of both the depth and displacement.

\begin{figure*}[!ht]\centering 
\includegraphics[width=0.98\linewidth]{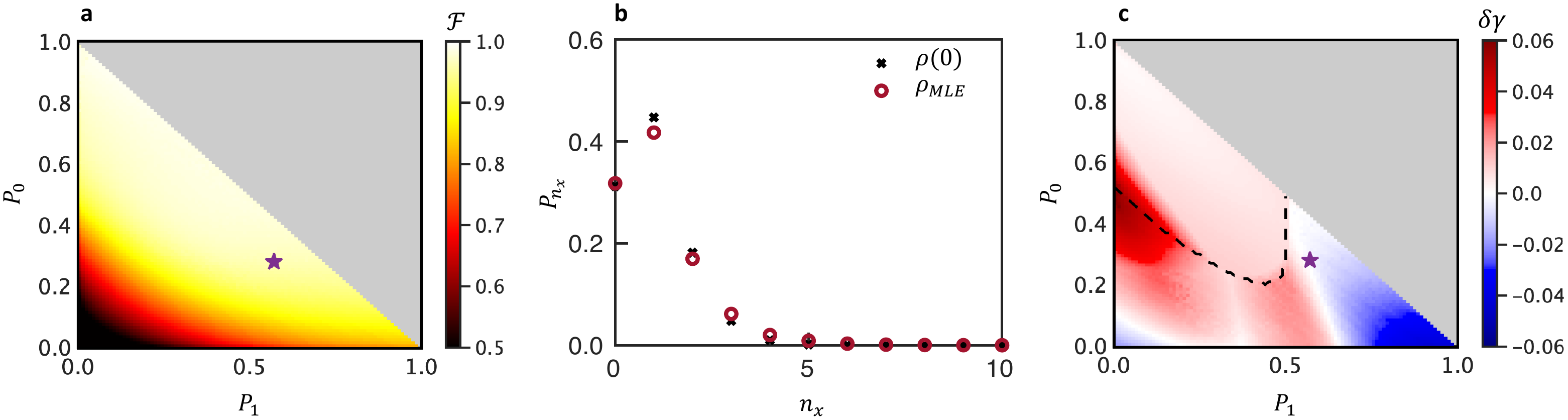}
\caption[Simulated Tomography Fidelity and Wigner Negativity]{
\textbf{(a)} Fidelity $\mathcal{F}$ (Eq.~\eqref{eq:fidelity}) between the state $\rho\left(0\right)$ and the reconstructed state $\rho_{\textrm{MLE}}$ for initial states $\rho_i$ characterized by a range of $\mathbf{P}=(P_{n=0}, P_{n=1}, P_{n=2})$, after evolution with our anharmonic model Eq.~\eqref{eq:H_quartic}. \textbf{(b)} Comparison of the diagonal elements for $n_x\leq 10$ of $\rho(0)$ and $\rho_{\textrm{MLE}}$ after a displacement back to the origin, corresponding to $\mathbf{P} = (0.28,0.57,0.15)$ (purple star in \textbf{a}). This point corresponds to a state with similar spatial extent as the $n_x=1$ state shown in Fig.~\ref{fig:Figure_3}e,h. Here, $n_x$ labels the Fock states of the oscillator in Eq.~\eqref{eq:H_quartic}. \textbf{(c)} For each $\mathbf{P}$, we then compute the Wigner negativity, $\gamma = \min_{x,p}W(x,p)$, for our initial state $\rho(0)$, as well as the Wigner negativity $\gamma_{\textrm{MLE}}$ of the reconstructed state. We plot the difference $\delta\gamma = \gamma - \gamma_{\textrm{MLE}}$; positive regions indicate parameters for which $\rho_{\textrm{MLE}}$ estimates more nonclassicality than the underlying state, whereas negative regions indicate parameters for which $\rho_{\textrm{MLE}}$ is an upper bound on this negativity. The dashed black dashed line separates nonclassical states with $\gamma < 0$ from those with $\gamma = 0$.}
\label{fig:supp_MLE_fid}
\end{figure*}

In Fig.~\ref{fig:supp_MLE_fid}a, we plot the fidelity of our reconstructed state, $\rho_{\mathrm{MLE}}$, to the actual underlying state, $\rho(0) = D(x_i)\rho_iD^\dagger(x_i)$, for a variety of compositions $\mathbf{P}$. In the figure, we indicate $\mathbf{P}$ that produces a state  before displacement with a similar spatial extent in the trap as the $n_x = 1$ states in Fig.~\ref{fig:Figure_3}e,h. This point, indicated by the purple star, corresponds to $\mathbf{P} = (0.28,0.57,0.15)$, and corresponds to estimates obtained in Fig.~\ref{fig:population_bar_plots} for the diagonal composition of the state prepared in the trap.

Within our model and for the indicated state, we find that the  reduction in the fidelity owing to anharmonic effects does not exceed 5\%. However, for higher populations of the $n_x=2$ state, a larger reduction in fidelity is expected from anharmonicity, and a deeper trap or smaller displacement would be needed to reduce the effect of anharmonicity and faithfully characterize such states.
Moreover, we find that the diagonal elements of the density matrix indicated by the purple star remain very similar (Fig.~\ref{fig:supp_MLE_fid}b). 

While the fidelity provides some measure of similarity between our states, of specific interest is the robustness of the nonclassical nature of the prepared states, which may be identified through the Wigner negativity $\gamma \leq 0$, where $\gamma = \min_{x,p} W(x,p)$. Therefore, we also plot the difference in negativity between $\rho(0)$ and $\rho_{\mathrm{MLE}}$ in Fig.~\ref{fig:supp_MLE_fid}c. Over the entire range of $\mathbf{P}$ considered, the difference in obtained negativity never exceeds $0.06$, and for the indicated estimates of our $n_x=1$ initial state composition, for which $\gamma \approx -0.04$, the corresponding MLE state remains negative with a difference in negativity of $< 0.01$.

\end{document}